\documentclass[article]{JHEP3} 

\usepackage{epsfig,multirow}
\usepackage{amsmath}
\usepackage{graphicx}
\usepackage[latin1]{inputenc}
\usepackage{amssymb}
\usepackage{amsmath}
\usepackage{verbatim}
\usepackage{epsfig}
\newcommand\fverb{\setbox\pippobox=\hbox\bgroup\verb}
\newcommand\fverbdo{\egroup\medskip\noindent%
                      \fbox{\unhbox\pippobox}\ }

\newcommand\fverbit{\egroup\item[\fbox{\unhbox\pippobox}]}
\newbox\pippobox
\catcode`~=11 
\newcommand{\urltilde}{\kern -.15em\lower .7ex\hbox{~}\kern .04em}
\catcode`~=13 
\newcommand{\ie}{{\it i.e.}}
\newcommand{\eg}{{\it e.g.}}

\newcommand{\pythia}{{\sc Pythia}}

\newcommand{\sherpa}{{\sc Sherpa}}
\newcommand{\mgme}{{\sc MadGraph/MadEvent}}
\newcommand{\alpgen}{{\sc Alpgen}}
\newcommand{\delphes}{{\sc Delphes}}
\newcommand{\pgs}{{\sc PGS}}
\newcommand{\madgraph}{{\sc MadGraph}}
\newcommand{\madevent}{{\sc MadEvent}}
\newcommand{\feynrules}{{\sc FeynRules}}
\newcommand{\python}{{\sc Python}}
\newcommand{\ufo}{{\sc UFO}}
\newcommand{\aloha}{{\sc ALOHA}}

\newcommand{\whizard}{{\sc Whizard}}
\newcommand{\helac}{{\sc Helac}}
\newcommand{\comix}{{\sc Comix}}
\newcommand{\cuttools}{{\sc CutTools}}
\newcommand{\madloop}{{\sc MadLoop}}
\newcommand{\madfks}{{\sc MadFKS}}
\newcommand{\powheg}{{\sc Powheg}}
\newcommand{\powhegbox}{{\sc PowhegBox}}
\newcommand{\comphep}{{\sc CompHEP/CalcHEP}}

\newcommand{\helas}{{HELAS}}
\newcommand{\mcatnlo}{{MC@NLO}}

\newcommand{\be}{\begin{equation}}
\newcommand{\ee}{\end{equation}}
\newcommand{\ba}{\begin{eqnarray}}
\newcommand{\ea}{\end{eqnarray}}
\newcommand{\bt}{\begin{tabular}}
\newcommand{\et}{\end{tabular}}
\newcommand{\bfig}{\begin{figure}}
\newcommand{\efig}{\end{figure}}

\preprint{}

\title{MadGraph 5 : Going Beyond} 

\author{Johan Alwall$^{(1)}$,  Michel Herquet$^{(2)}$,   Fabio Maltoni$^{(3)}$,  Olivier Mattelaer$^{(3)}$,  Tim Stelzer$^{(4)}$\\
$^{(1)}$ Theoretical Physics Department, Fermi National Accelerator Laboratory,
P. O. Box 500, Batavia, IL 60510, USA\\
$^{(2)}$ Nikhef Theory Group, Kruislaan 409, 1098 SJ Amsterdam, The Netherlands\\
$^{(3)}$ Centre for Cosmology, Particle Physics and Phenomenology (CP3) 
Universit\'{e} Catholique de Louvain,
Chemin du Cyclotron 2, B-1348 Louvain-la-Neuve, Belgium\\
$^{(4)}$ Department of Physics, University of Illinois at Urbana-Champaign, 
1110 West Green Street, Urbana, IL  61801 
}

\abstract{\madgraph~5 is the new version of the \madgraph\ matrix
element generator, written in the Python programming language. 
It implements a number of new, efficient algorithms that provide
improved performance and functionality in all aspects of the
program. It features a new user interface, 
several new output formats including C++ process libraries for \pythia\ 8, and 
full compatibility with \feynrules\ for new physics models
implementation, allowing for event generation for any model that can
be written in the form of a Lagrangian. \madgraph~5 builds on the same
philosophy as the previous versions, and its design allows it to be used as a  collaborative platform where
theoretical, phenomenological and simulation projects can be developed and then distributed 
to the high-energy community. We describe the ideas and the most
important developments of the code and illustrate its capabilities through a 
few simple phenomenological examples.}

\date{\Date}

\keywords{Monte Carlo, MadGraph, Hadronic Colliders}

\begin{document}

\section{Introduction}


Identifying the fundamental building blocks of matter and describing
their interactions from first principles is the goal not only of
current accelerator based experiments, such as those operating at
Tevatron and the LHC, but also of many other experiments, including
flavour and neutrino experiments, and dark matter detection experiments
in underground laboratories or satellites.

Discoveries from these experiments as well as their interpretation will rely on our ability
to perform accurate simulations for both the signals and their backgrounds.  At the LHC, for instance,
extracting physics from the data will present several significant challenges. 
First, proton-proton collisions at very high energies produce final states that involve a large number of jets, heavy-flavour quarks, leptons and missing energy, providing an overwhelming 
background to many new physics searches. Second, even in presence of a clear ``anomaly"
with respect to the Standard Model prediction, its interpretation in terms of an underlying
phenomena or theory could be extremely difficult. Tools that are able to make
precise predictions for wide classes of Beyond the Standard Model (BSM) physics, as well
as those that help in building up an effective field theory from the
data, will be employed.  Before one or a few candidate theories can be selected,
accurate measurements of the corresponding parameters (masses, couplings, spin, charges) will be
needed. Production rates and/or branching ratio measurements, for example, 
will provide constraints only if we are able to connect them  to the fundamental parameters
of a model through an accurate calculation, at least at next-to-leading order in perturbative QCD. 

In this context, there is no doubt that Monte Carlo simulations  play a key role at each stage of the exploration of the TeV scale, \ie, from the discovery and identification of BSM physics,
to the measurement of its properties.  The realization of the need
for better simulation tools for the LHC has spurred an intense
activity in recent years, that has resulted in several important advances in the field.

General purpose matrix-element based event gene\-ra\-tors, such as 
\comphep~\cite{Pukhov:1999gg,Boos:2004kh,Pukhov:2004ca},
{\sc MadGraph/ MadEvent}~\cite{Stelzer:1994ta,Maltoni:2002qb,Alwall:2007st}, 
\sherpa~\cite{Gleisberg:2003xi} and 
\whizard~\cite{Kilian:2001qz} have been available for several years now. More recently, highly efficient multiparton techniques which go beyond 
usual Feynman diagrams have been introduced~\cite{Caravaglios:1995cd,Draggiotis:1998gr,Duhr:2006iq}, 
and implemented in publicly available codes, such as \alpgen~\cite{Mangano:2002ea}, \helac~\cite{Papadopoulos:2006mh} and \comix~\cite{Gleisberg:2008fv}. 
As a result, the problem of automatically generating tree-level matrix elements (and then cross
sections and events) for a very large class of renormalizable
models has been solved. The recent introduction of \feynrules~\cite{Christensen:2008py} has provided a new  method for  implementing new physics models as well as setting a new standard in terms of validation and availability~\cite{Christensen:2009jx,Duhr:2011se}.
Communication between \feynrules\  and matrix element programs is being standardized via the new Universal \feynrules\ Output format, the \ufo~\cite{ufo:2011}.

A connected effort is being made in the automation of NLO computations. The generation of the real corrections with the appropriate subtractions has been achieved in an automatic
way~\cite{Gleisberg:2007md,Seymour:2008mu,Hasegawa:2008ae,Frederix:2008hu,Czakon:2009ss,Frederix:2009yq}. For
virtual corrections, several new algorithms for numerical calculation of loop amplitudes
have been proposed (see, \eg, \cite{Zanderighi:2008na} for a review)
and some of them successfully applied to the computation of SM
processes of physical interest~\cite{Ellis:2009zw,Berger:2009zg,vanHameren:2009dr,Berger:2010vm,Berger:2010zx}.
Very recently, \cuttools~\cite{Ossola:2007ax} has been successfully
interfaced with \madgraph. The resulting tool, \madloop\ \cite{Hirschi:2011pa} interfaced to \madfks\ \cite{Frederix:2009yq},  allows a fully automatic calculation of  infrared-safe observables at NLO in QCD for a wide range of processes in the Standard Model. 

Last but not least, an accurate simulation of a hadronic collision requires a careful
integration of the matrix-element hard process, with the full parton
showering and hadronization infrastructure \cite{Sjostrand:2006za,Corcella:2000bw,Gleisberg:2008ta}. 
Here again, significant progress has been made in the development of
merging algorithms, such as CKKW and MLM merging \cite{Catani:2001cc,Krauss:2002up,Mrenna:2003if,Mangano:2006rw,Lonnblad:2001iq,Lavesson:2005xu,Hoeche:2009rj}, and  in their
comparison~\cite{Hoche:2006ph,Alwall:2007fs}, with applications to
SM~\cite{Krauss:2004bs,Mangano:2006rw,Englert:2011cg} and to BSM~\cite{Alwall:2008qv}
processes.  A breakthrough in merging fixed order calculations and parton showers
was achieved in Refs.~\cite{Frixione:2002ik,Frixione:2003ei}, where it is shown how to
correctly interface an NLO computation  to avoid
double counting and delivered the first event generator at NLO,
\mcatnlo. More recently, another method along the same lines, dubbed
\powheg, has been proposed~\cite{Nason:2004rx} and applied to a variety of
processes at the LHC through the \powhegbox\ general implementation~\cite{Alioli:2010xd}.

The new version of \madgraph\ has been designed to support and advance the lines of
development mentioned above,  with three main objectives:

\begin{enumerate}
\item Lagrangian-based BSM physics via \feynrules\ for any renormalizable or effective theory.
\item Full automation and optimization of NLO computations in the SM and beyond.
\item Merging to showering/hadronization codes for complete event simulation at 
LO (via CKKW and MLM methods) and at NLO (via \mcatnlo\ and \powheg), as
well as the combination of the two (``CKKW at NLO'').
\end{enumerate}

\madgraph~5 is open source software written in Python and features a
collaborative development structure. It can generate matrix elements
at the tree-level for any  Lagrangian based model (renormalizable or effective)
implemented in \feynrules\ via the \ufo\ interface, and automatic
generation of the corresponding helicity amplitude
subroutines via the \aloha\ package~\cite{aloha:2011}. With respect to
\madgraph~4, significant efficiency improvement has been attained, and the
possibilities for tree-level matrix element generation (and diagram
plotting) have been extended, including optimization of the \madevent\ output 
and reorganization of multi-jet final state subprocesses.  
It features a wide set of flexible output formats in Fortran, C++, and
Python, and dedicated matrix element output for \pythia~8 \cite{Sjostrand:2007gs}.

This work documents and describes the general philosophy of the new
\madgraph\ version, as well as some of the most important improvements
in the code. The paper is structured as follows: In
Sec.~\ref{sec:overview} we give a general overview on the code
structure and of the algorithms employed.  In dedicated subsections we
describe the diagram generation algorithm, the fermion-flow algorithm, the colour algebra module,
and the generation of decay chains. We then present the available output formats, the new multiprocesses
optimization in \madevent\ and process library generation for
\pythia~8 as well as the new diagram drawing algorithm. Model inheritance from \feynrules\ via the \ufo\ and \aloha\ is presented  in Sec.~\ref{sec:models} together with a comprehensive list of available models and the
suite of model tests that can be performed at the process level.
The following section describes the validation
checks that have been performed for SM, MSSM, HEFT and RS processes, as well as
some key indicators for the performance improvements in speed compared to previous versions of \mgme.
Section~\ref{sec:examples} provides a selective set of examples of
applications. We leave our conclusions and the discussion of the
outlook to the last section. Technical appendices follow, where
the interested reader can find more details and examples.

\section{Overview and algorithms}
\label{sec:overview}

\madgraph\ \cite{Stelzer:1994ta} is a tool for automatically
generating  matrix elements for High Energy Physics processes, such as
decays and $2 \to n$ scatterings. First, the user specifies a process in terms
of initial and final state particles (allowing for a number of refined
criteria, including forced or forbidden $s$-channel resonances, excluded internal
particles, and forced decay chains of final state
particles). Multiparticle labels can be used to specify all possible
processes involving a range of particles. As a result, \madgraph\  generates
all Feynman diagrams for the process, and outputs the computer code
necessary to evaluate the matrix element at a given phase space
point. The matrix element evaluation is done using calls to helicity
wavefunctions and amplitudes, as were first implemented in the \helas\
package \cite{Murayama:1992gi}. This implementation is efficient
 because it naturally allows helicity wavefunctions corresponding to identical subdiagrams
to be reused across diagrams. \madgraph\ also produces
pictorial output of the Feynman diagrams for the process in question.
The computer code produced by \madgraph\ can then be used for cross
section or decay width calculations and event generation, e.g.\ using
the \madevent\ package \cite{Maltoni:2002qb}, which is included with
\madgraph~5.

While previous versions of \madgraph\ were written in Fortran 77,
\madgraph~5 is written in Python. This object oriented computer language allows for completely new
algorithms, and removes may of the restrictions that were inherent in the Fortran
versions. As a result, both \madgraph~5 and the code it produces run significantly
faster than previous versions. Even more important
however, is that the structure of the new implementation greatly
facilitates selective use of modules and additions of new features. In this paper we will discuss a
few such additions, such
as implementation of new colour representations (colour sextets and
$\epsilon^{ijk}$), implementation of multi-fermion vertices and
addition of new output formats, including output in C++ and Python.

\subsection{Diagram generation}
\label{sec:diagram_generation}

The diagram generation algorithm used in \madgraph~5 is faster, more efficient
and produces better optimized code than earlier \madgraph\ versions. 

In \madgraph~4 diagrams generation is based on the following algorithm:
\begin{enumerate}
\item Generate all topologies with appropriate number of external legs.
\item Assign particles to external legs.
\item Identify vertices with at most one unassigned line.
\item Check to see if there is any interaction in the model that will
  accomodate the assigned lines.
\item If yes, 
\begin {enumerate} 
\item if a vertex had an unassigned line, assign to it the appropriate
  particle ID from the interaction, then check next vertex.
\item if all lines in the vertex are known, diagram is complete. Write it to file.
\end{enumerate}
\item If no, diagram fails, try next topology.
\end{enumerate}

This algorithm is straightforward to implement, but the time requirement grows quickly with 
the number of external particles since every topology must be explicitly checked, even if only a 
small fraction contribute to viable diagrams.   

The algorithm in \madgraph~5 eliminates this inefficiency by making use of the model information to effectively
only construct topologies that will yield a valid diagrams.  Furthermore, it
recursively generates all of the diagrams in parallel. This ensures that any
combination of external legs (a,b), (a,b,c) etc.\ that is common to
multiple diagrams will be recognized as such,  allowing for optimal
recycling of already calculated subdiagrams in the resulting helicity
amplitude code. It also removes restrictions on the number of
particles in an interaction vertex, paving the way for 
implementation of higher-dimensional effective interactions with 
5 or more fields.

The algorithm, presented below, is based on recursively creating
sub-diagrams from the diagrams by merging legs,
with the crucial addition of a flag, \texttt{from\_group}, which is used to indicate whether a given
particle results from a merging of particles (\ie, it is connected to a given set of particles) in the previous step
(\texttt{True}), or if it is simply copied from the previous step
(\texttt{False}). This flag helps ensure that no diagrams are
double-counted by the algorithm.

\begin{enumerate}
\item Given the model, generate two hash maps (called dictionaries in
 Python), containing information about the interactions in
 the model. The first dictionary (called \texttt{Vertices}) maps all
 combinations of $n$ particles
 to all $n$-point interactions combining these particles, and maps all
 pairs particle-antiparticle to ``\texttt{0}''. The second dictionary
 (called \texttt{Currents}) maps, for all $n$-point interactions, $n-1$ particles to all combinations of
 resulting particles for the interactions.
\item Flip particle/anti particle status for incoming particles in the
  process (\ie, consider all the particles outgoing). Set the flag
  \texttt{from\_group} = \texttt{True} for all external particles.
\item If there is an entry in the   \texttt{Vertices} dictionary combining all external particles,
create the combination [(1,2,3,4,...)] if \textbf{at least two}
particles have \texttt{from\_group} = \texttt{True}.
\item Create all allowed groupings of particles with \textbf{at least
one} \texttt{from\_group}=\texttt{True} present in the \texttt{Currents}  dictionary.
\item Set \texttt{from\_group}=\texttt{True} for the newly combined
particles, and \texttt{False} for any particle that has not been
combined in this iteration. Repeat from \textbf{3} for the reduced set
of external particles.
\item Stop algorithm when at most 2 external particles remain.
\end{enumerate}

As a simple, yet complete example, let us consider the process $e^+e^-\to u\bar u g$ in the standard model.
The procedure is illustrated in Table.~\ref{tab:epemuubarg_diagrams},
and described in detail below.
The relevant interactions are  $(e^+e^-\gamma)$, $(e^+e^-Z)$, $(u\bar u\gamma)$, 
$(u\bar uZ)$, and $(u\bar ug)$. 

\begin{table*}
\centering
\begin{tabular}{|c|c|c|c|}
\hline
 1st iteration & Groupings & After replacements & Result \\
\hline
\multirow{17}{*}{$e^-,e^+, u, \bar u, g$} &
\multirow{2}{*}{$(e^-,e^+),u,\bar u,g$} & $(\gamma), u, \bar u, g$ &
Failed (only 1 FG=\texttt{True}) \\
\cline{3-4}
&& $(Z), u, \bar u, g$ & Failed (only 1 FG=\texttt{True}) \\
\cline{2-4}
& \multirow{3}{*}{$e^-,e^+,(u,\bar u),g$} & $e^-,e^+,(\gamma),g$ &
Failed (only 1 FG=\texttt{True}) \\
\cline{3-4}
& & $e^-,e^+,(Z),g$ &
Failed (only 1 FG=\texttt{True}) \\
\cline{3-4}
& & $e^-,e^+,(g),g$ &
Failed (only 1 FG=\texttt{True}) \\
\cline{2-4}
& $e^-,e^+,(u,g),\bar u$ & $e^-,e^+,(u),\bar u$ &
Failed (only 1 FG=\texttt{True}) \\
\cline{2-4}
& $e^-,e^+,u,(\bar u,g)$ & $e^-,e^+,u,(\bar u)$ &
Failed (only 1 FG=\texttt{True}) \\
\cline{2-4}
& \multirow{6}{*}{$(e^-,e^+),(u,\bar u),g$} & $(\gamma),(\gamma),g$ &
Failed (no vertex) \\
\cline{3-4}
& & $(\gamma),(Z),g$ &
Failed (no vertex) \\
\cline{3-4}
& & $(\gamma),(g),g$ &
Failed (no vertex) \\
\cline{3-4}
& & $(Z),(\gamma),g$ &
Failed (no vertex) \\
\cline{3-4}
& & $(Z),(Z),g$ &
Failed (no vertex) \\
\cline{3-4}
& & $(Z),(g),g$ &
Failed (no vertex) \\
\cline{2-4}
& \multirow{2}{*}{$(e^-,e^+),(u,g),\bar u$} & $(\gamma), (u), \bar u$ &
Diagram 1\\
\cline{3-4}
&& $(Z), (u), \bar u$ & Diagram 2 \\
\cline{2-4}
&\multirow{2}{*}{$(e^-,e^+),u,(\bar u, g)$} & $(\gamma), u, (\bar u)$ &
Diagram 3\\
\cline{3-4}
&& $(Z), u, (\bar u)$ & Diagram 4 \\
\hline
\end{tabular}
\caption{\label{tab:epemuubarg_diagrams}
Tabel to illustrate the steps of the diagram generation algorithm. See
text for explanations. \texttt{from\_group} has been abbreviated as FG.}
\end{table*}

First iteration:

After flipping the particle/antiparticle identities
for the initial state, we have the external particles $e^-,e^+, u, \bar u, g$.
\begin{enumerate}
\item No grouping $(e^-,e^+, u, \bar u, g)$ is possible.
\item Create all possible particle groupings (see Table~\ref{tab:epemuubarg_diagrams}):\\
$[(e^-,e^+), u, \bar u, g]$, 
$[e^-,e^+,(u, g), \bar u]$, 
$[e^-,e^+,u,(\bar u, g)]$,\\
$[(e^-,e^+),(u,\bar u), g]$, 
$[(e^-,e^+),(u, g), \bar u]$, 
$[(e^-,e^+),u,(\bar u, g)]$\\ 
and replace the grouped particles with the resulting particles from the interactions:\\
$[(\gamma), u, \bar u, g]$, 
$[(Z), u, \bar u, g]$, 
$[e^-,e^+, (\gamma), g]$, 
$[e^-,e^+, (Z), g]$,  
$[e^-,e^+, (g), g]$,\\
$[e^-,e^+, (u), \bar u]$,   
$[e^-,e^+, u, (\bar u)]$,
$[(\gamma), (\gamma), g]$,   
$[(Z), (\gamma), g]$, 
$[(\gamma), (Z), g]$, 
$[(Z), (Z), g]$,\\
$[(\gamma), (g), g]$,
$[(Z), (g), g]$,
$[(\gamma), (u), \bar u]$, 
$[(Z), (u), \bar u]$,
$[(\gamma), u, (\bar u)]$,
$[(Z), u, (\bar u)]$. 
Note that only the particles in
parentheses now have \texttt{from\_group} = \texttt{True}.
\end{enumerate}

Second iteration:

The resulting reduced sets with four particles all have only one \texttt{from\_group}
= \texttt{True}, and can therefore not give valid diagrams. We
therefore ignore these and focus on the reduced sets with three particles.
\begin{enumerate}
\setcounter{enumi}{2}
\item The combinations allowed by the interactions are:\\
$((\gamma), (u), \bar u)$, $((Z), (u), \bar u)$, 
$((\gamma), u, (\bar u))$ and $((Z), u, (\bar u))$.
\item Any further combination in the \texttt{Currents} dictionary will result in an external
state with only one \texttt{from\_group} = \texttt{True}, which can not
give any diagrams.
\item The iteration stops, since no external particles are left.
\end{enumerate}

\begin{figure}
\centering
\includegraphics[trim=70 320 -120 -280,width=12cm,clip=true]{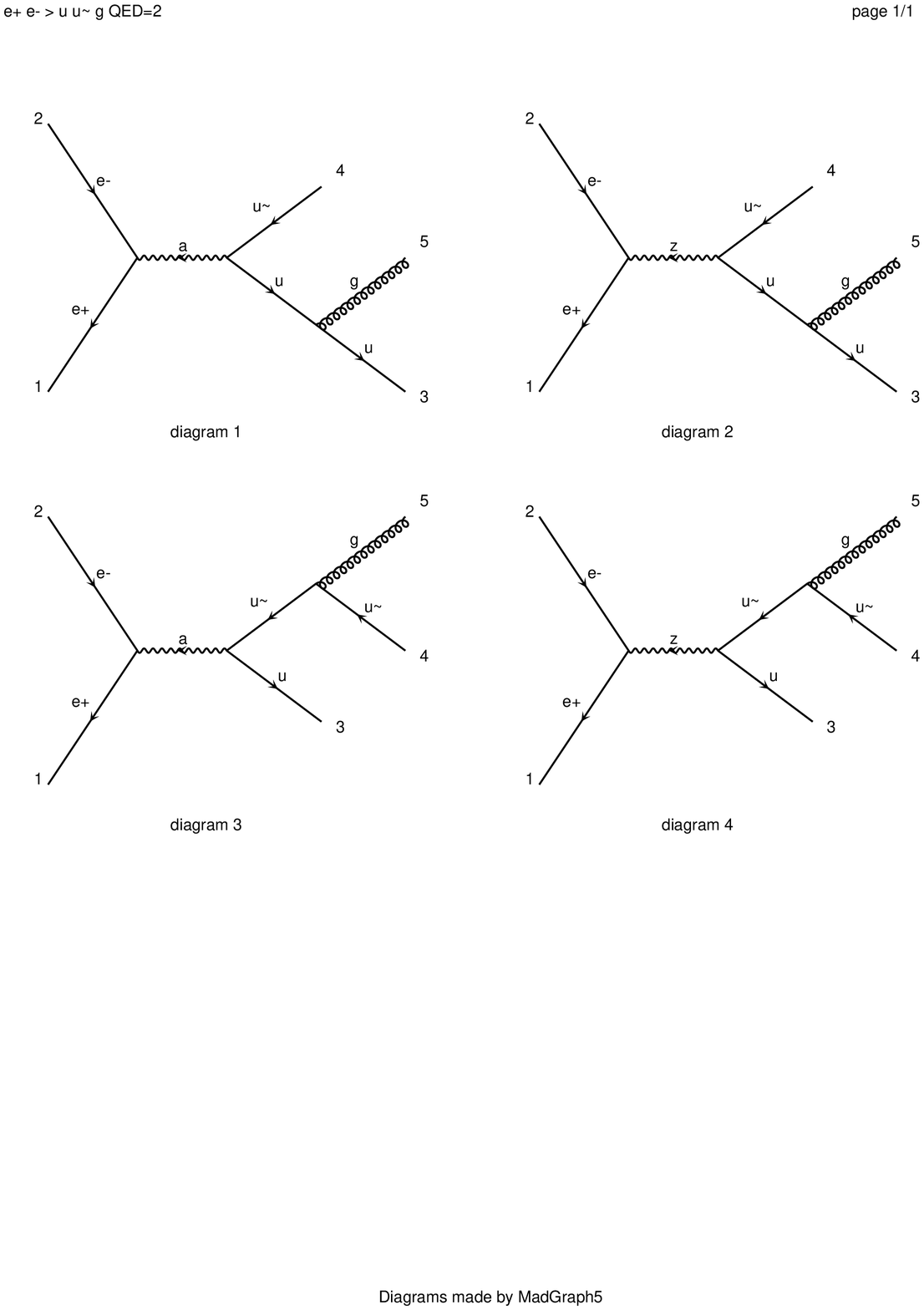}
\caption{Diagrams for the process $e^+e^-\to u\bar u g$. Note that
  $\bar u$ is denoted by ``u\urltilde'' and $\gamma$ by ``a'' in the diagrams.}
\label{fig:epemuubarg_diagrams}
\end{figure}

The resulting diagrams are found in Fig.~\ref{fig:epemuubarg_diagrams}.

\subsection{Helicity amplitude call generation and fermion number violation}
\label{sec:fermionflows}

The code for matrix element evaluation generated by \madgraph\
(previous versions as well as \madgraph~5) is written in terms of successive
calls to a helicity amplitude function library (originally the \helas\
library, in \madgraph~5 either \helas\ or helicity
amplitude functions automatically generated by \aloha~\cite{aloha:2011}). A helicity wavefunction is
generated for each external leg in a diagram, and these wavefunctions
are combined into new wavefunctions corresponding to the propagators in
the diagram by successive helicity wavefunction calls. The final
vertex corresponds to a helicity amplitude call which returns the
value of the amplitude corresponding to this diagram. In this
procedure, wavefunctions corresponding to identical subdiagrams
contributing to different diagrams can be reused between the diagrams,
leading to a considerable optimization, effectively giving up to a
factor hundred fewer wavefunction calls than naively expected by the number of
Feynman diagrams.

In the presence of Majorana particles or fermion number
violating vertices, special care is needed to allow for all
possible contractions of fermions. \madgraph~5, like its predecessors
\cite{Cho:2006sx,Alwall:2007st}, uses the 4-spinor Feynman rules for
fermion number violation developed
in \cite{Denner:1992vza}. In this formulation, a fermion flow is
defined for each fermion line in a diagram, and fermion number
violation (due to fermion flow clashes) is taken into account using special charge
conjugate versions of the $\Gamma^\mu$ matrices. 
With this modification, the fermion lines meeting at
the vertex should be treated as if they had a single fermion flow. The
flow therefore needs to be inverted along one of the lines, 
resulting in a continuous fermion flow.

In earlier versions of \madgraph, this fermion flow was implemented by
creating a charge-conjugate particle for each fermion in the model. In
order to check for diagrams with clashing arrows it was necessary to check
each topology with both the regular fermion, and its charge conjugate. This
would increase the time for generating code by a factor of $2^{N_{\rm fermions}-1}$.

In \madgraph~5, the diagrams resulting from the diagram generation are
independent of the fermion flow, and the definition of the flow is
postponed to the time of helicity amplitude call generation. 

Fermions with positive PDG code (``particles'') are tentatively assigned to be
incoming (outgoing) if they are in the initial (final) state, and vice
versa for fermions with negative PDG code (``antiparticles''). If a
fermion flow clash is detected at the meeting of two fermion lines
(\ie, the two fermions have the same ``incoming/outgoing'' status),
the fermion flow direction of one of the lines has to be inverted, and
charge conjugate vertices have to be introduced starting from the position of the
vertex or Majorana particle line responsible for the fermion flow clash.

The procedure is the following: Fermion lines involved in the clash
are traversed, looking for Majorana particles. If a Majorana particle
is found along one of the lines, the ``incoming/outgoing'' status and
particle/antiparticle id is reversed for all particles up to (and
including) the last Majorana particle along the line. For particles
beyond the Majorana particle along the same fermion flow line, a flag
\texttt{fermionflow} is set to -1. The other line is left unchanged.
If no Majorana particle is found along either of the lines (which is
the case when the clash is due to a fermion flow violating vertex),
all fermions along the first leg have their \texttt{fermionflow} flag
is set to -1.

A helicity amplitude or wavefunction with any fermion with negative
\texttt{fermionflow} flag is required to use the
charge conjugate version of the amplitude of wavefunction, in
accordance to the Feynman rules in \cite{Denner:1992vza}. For an
external leg, the incoming/outgoing status is reversed. 

For multifermion vertices, the fermions are grouped in pairs, each
constituting a fermion line (see Sec.~\ref{sec:4fermion}). Each
fermion line then has its own charge conjugate, which can lead to
multiple charge conjugate flags for a single helicity amplitude or
wavefunction.

\begin{figure}
\centering
\includegraphics[trim=70 570 160 -280,width=6cm,clip=true]{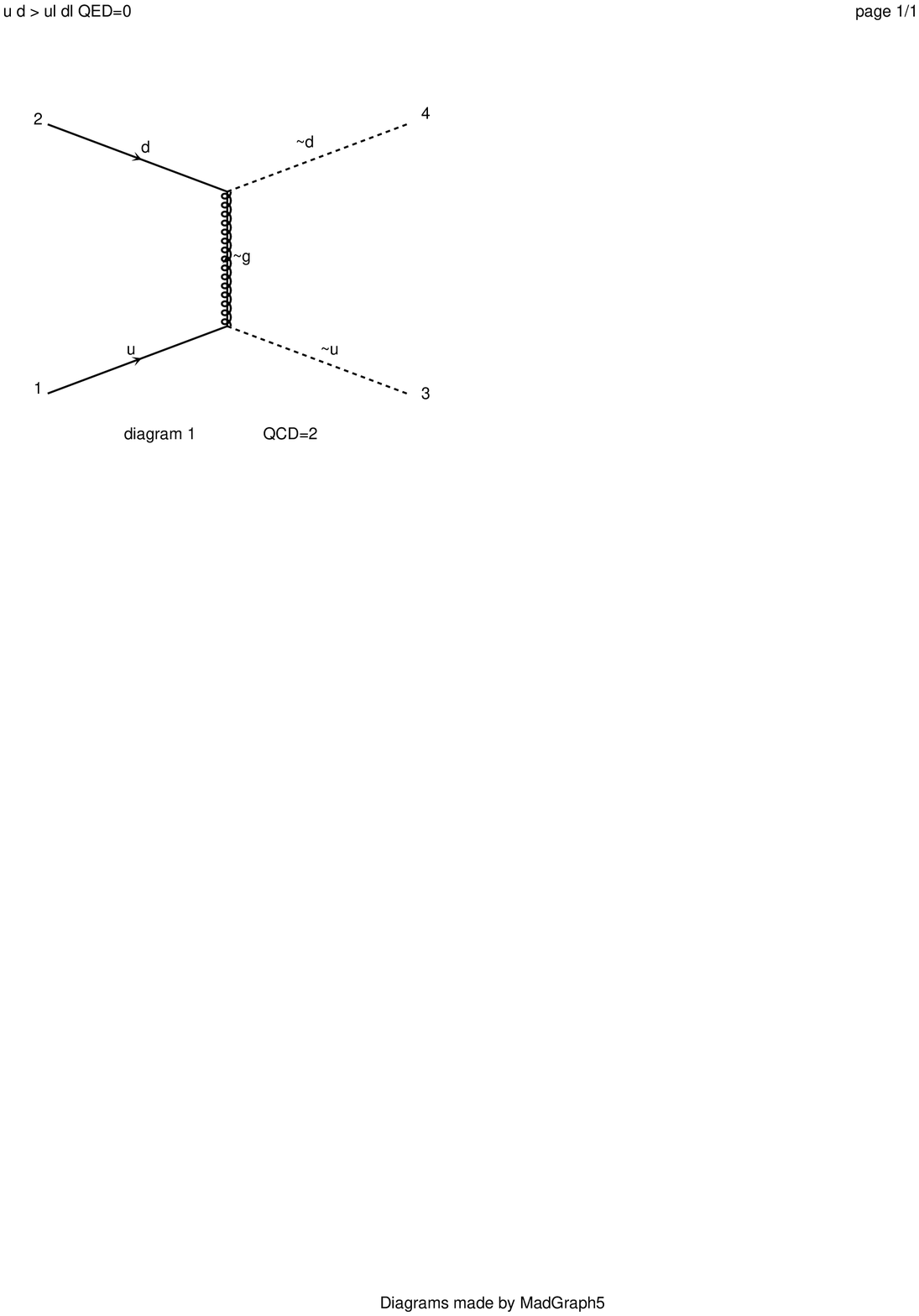}
\caption{Diagram for the fermion flow violating process $u d \to
  \tilde u \tilde d$.}
\label{fig:ud_susd_diagram}
\end{figure}

As an explicit example of both helicity amplitude call generation and
the treatment of fermion flow violation, we take the fermion number
violating process $u d \to \tilde u\tilde d$ with $t$-channel exchange
of a Majorana gluino (see Fig.~\ref{fig:ud_susd_diagram}).  The
diagram is represented internally (with tentative incoming/outgoing
fermion state given in parentheses) as
$$
(u (\text{in}),\tilde u \to \tilde g (\text{in})), 
(\tilde g (\text{in}), d (\text{in}), \tilde d)
$$
The $u\tilde u\to \tilde g$ vertex would result in an incoming spin
$\frac12$ helicity wavefunction with incoming flow, and the $\tilde g
d \tilde d$ vertex correspond to a helicity amplitude call. However, a
fermion flow clash is detected in the last vertex, since there are two
incoming fermions. The two fermion lines coming from this vertex
(initiated by the $\tilde g$ and $d$ respectively) are now traversed,
starting with the gluino. Since the gluino is a Majorana particle (and
there are no other Majorana particles along the line), its flow is
reversed to ``outgoing''. The $u$, which is beyond the gluino along
the line, gets the flag \texttt{fermionflow} set to -1. This indicates
that the $u$ wavefunction should be that of an outgoing external spin
$\frac 12$ particle, and the $u \tilde u \to \tilde g$ wavefunction
should be conjugated (since one of the particles in this wavefunction
has \texttt{fermionflow} flag -1). The $\tilde g d \tilde d$ helicity
amplitude does not use the conjugated version, since none of the
involved wavefunctions have negative \texttt{fermionflow} flag.

\subsection{Colour algebra}
\label{sec:colour_algebra}

In this section we discuss how the  computation of colour coefficients in scattering amplitudes of states which are charged under $SU(3)_C$ is performed. There are two main motivations to dedicate special
attention to this aspect.

First, multiparton amplitudes, \ie, amplitudes with many external quarks and gluons, are 
phenomenologically very  important as they correspond to the leading order 
approximation of multi-jet production (by themselves or in association with other
particles) in high energy collisions and especially at hadron colliders.
These processes are the major backgrounds to many
new-physics signals, so an accurate description of these final states is
essential. In this case, not only the complexity of colour computation  grows factorially 
but also the amount of information to be stored grows at the same rate. Efficiency and
improved algorithms are therefore necessary.

Second, new-physics models possibly feature states in exotic colour representations
or non-standard colour structures in interaction vertices, which need to be taken
into account. A generic interaction vertex may also have a complicated 
structure with several colour factors in front of different Lorentz structures. 

The needs above require both flexibility and efficiency of the colour algebra module,
a challenge that is not easy to meet.

Prior versions of \madgraph\ hard-coded three colour structures $\delta_{ij}$, $T^{a}_{ij}$ and $f^{abc}$
as well as identities for summing over the colour indices.  The number of colours was explicitly set to 3
in these identities. Numeric values for the colour matrix where computed by summing over the colour indices using integer operations on the numerator and denominator separately.  The resulting colour matrix was exact, but lacked flexibility. The colour structure of the interaction was inferred based on the colour of the participating particles. Models that had new colour interactions
required the user to explicitly code new colour structures, which required detailed knowledge of the \madgraph\ code,
significantly restricting the types of new interactions that could be implemented.

The solution adopted in \madgraph~5
is to have all the colour objects and their algebra coded in a symbolic way.
This approach has many advantages, the most important one being
complete flexibility.

For example, one important aspect in having efficient colour computations is that 
of the choice of the colour basis.
Several possibilities have been proposed in the literature 
\cite{Mangano:1990by,DelDuca:1999rs,Maltoni:2002mq} with the aim  to make the computation
of colour symbolically and/or numerically efficient at the amplitude level. All these choices
can be easily adopted in our implementation as  the 
colour algebra is dealt with symbolically and several different basis and representations
can be present at the same time and used as needed. 
The colour algebra is realized through objects, such as the Gell-Mann matrices or structure
functions, whose products and combinations can be easily simplified by 
algebraic reduction rules, among which
\begin{equation}
(T^{a})^i_{j} (T^{a})^k_{l} = \frac12 \left( \delta^i_{l} \delta^k_{j} - \frac{1}{N_c}   \delta^i_{j} \delta^k_{k} \right)
\end{equation}
plays a central role. Other objects, such as, $\delta^{ab}$, $f^{abc}$ and $d^{abc}$ 
are defined in terms of linear combinations of traces of Gell-Mann matrices, and any colour factor
can be easily simplified in a recursive way.  

Our default algorithm goes as follows. The colour factor corresponding to each diagram
is linearly decomposed over a complete and orthogonal (in the large $N_c$ limit) basis which is constructed
at the same time. This allows the automatic organization of the full amplitude
into gauge invariant subamplitudes ${\cal A}_i$ (normally called dual or colour-ordered amplitudes),
each de facto corresponding to a given colour flow $i$, so that 
\begin{equation}
{\cal M} = \sum_{i} {\cal C}_i {\cal A}_i \,.
\label{colour_decomp}
\end{equation}
The form above is the basis for further manipulations, as it can be now used in different ways
depending on the complexity of the calculation itself. In the case of a limited number of
partons in the amplitude, it is squared by analytically computing 
the colour matrix ${ \cal C}_{ij} = \sum_{\rm colours}   { \cal C}_{i}  { \cal C}^*_{j}$, which
is then stored in memory and written in the output file. This is the default approach 
followed in this version of \madgraph~5 where all diagrams are computed and the colour ordered amplitudes
obtained. At this point we note that 
in the presence of identical external particles, typically gluons, many of the entries of the colour matrix are equal due to simple symmetry properties. Such symmetries are efficiently exploited to reduce the effective
number of computations needed to determine the full ${ \cal C}_{ij}$. This allows to calculate  amplitudes with up to 
7 gluons. Beyond 7 gluons, explicitly calculating the amplitude associated with each Feynman diagram is not viable and recursive relations need to be employed \cite{Berends:1987me,Britto:2004ap}. These allow one to calculate the ${\cal A}_i$ directly and prove that their complexity is only
polynomial. The problem of the factorial growth therefore remains only for the colour sum.
Several techniques have been proposed, the most commonly employed is the
idea of recursively computing ${\cal M}$ at fixed colour for the external states and then randomly choose
colour configurations \cite{Mangano:2002ea,Papadopoulos:2006mh,Duhr:2006iq}. 
Another possibility, which is born out of Eq.~\ref{colour_decomp}, is to organize the calculation as an expansion in $1/N_c$. 
This is the approach that is currently under investigation to calculate multiparton amplitudes at the tree and loop level and
also to combine real and virtual corrections in NLO computations, following the approach of Ref.~\cite{Frixione:2011kh}.

The other main advantage of treating the colour algebra symbolically is a  straightforward implementation of new colour structures. As an example, we here give a detailed
description of the necessary algebra for colour sextets and colour
triplet $\epsilon$ tensors. Higher dimensionality colour representations can
be implemented in a similar way.

The $\epsilon^{ijk}(\bar\epsilon_{ijk})$ object is the totally antisymmetric tensor of
three colour triplet (antitriplet) indices. The algebraic relations
needed for $\epsilon$ and $\bar\epsilon$ are:
\begin{equation}
\epsilon^{ijk}\bar\epsilon_{ilm} = \delta^j_l \delta^k_m - \delta^j_m \delta^k_l \nonumber
\end{equation}
where 
$(\epsilon^{ijk})^* = \bar\epsilon_{ijk}$. For colour sextets, we need three new colour objects $(K_6)^A_{ij}$,
$(\overline{K_6})_A^{ij}$ and the sextet representation $(T_6^{a})^A_B$. $K_6$ ($\overline{K_6}$) is the
symmetric tensor contracting a colour sextet (antisextet) index and two
colour antitriplet (triplet) indices (\ie, the Clebsch-Gordan coefficient), while the $T_6$ 
describes the interaction of a gluon with a sextet. As in the colour triplet implementation
described above, the sextet delta function
is denoted as a two-index $\delta^m_n$. The complete set of needed
algebraic relations for these objects can be found in the Appendix of Ref.~\cite{Han:2009ya}.
They can be easily reproduced by starting from
\begin{equation}
(K_6)^A_{ij}(\overline{K_6})_A^{kl} = \frac12(\delta^l_i \delta^k_j + \delta^k_i \delta^l_j). \nonumber
\end{equation}
Together with fundamental anti-commutation relation, 
the colour matrix and colour flows of any diagram involving colour sextet particles can be calculated.

As a final remark, we note that a small technical complication arises if
the parton level events are passed to a parton shower.  In this case 
the writing of the event in the Les Houches Accord (LHA) at leading-$N_C$ colour strings
is needed. The $\epsilon$ tensor can be handled by inserting new colour labels in the event
and even though an apparent violation of the colour flow arises, this can be interpreted
correctly by the parton-shower (see Ref.~\cite{Alwall:2006yp}). For colour sextets no convention
has been established. A simple solution is to note that in 
the planar ``double line'' colour flow notation, a colour sextet is contracted with a 
$K_6$, to give the equivalent pair of triplet lines. 
This can then be written into the LHA event, by using a negative antitriplet label
for the second triplet, and vice versa for a second antitriplet. A
parton shower program reading the event must then treat the continued
colour flow, keeping track of colour sextets and antisextets
appropriately.

\subsection{Decay chains}
\label{sec:decay}

Whereas decay chains in \madgraph~4 were generated in the same way as
regular process generation (by dressing topologies with particle and
interaction information to create diagrams),
decay chains in \madgraph~5 are defined using successive chains of
processes. The core process can have a number of decay processes
defined, and each decay process can again have a number of decay
processes defined, and so on. This treatment allows for quick and
efficient generation of decay chains of virtually unlimited length.

The helicity amplitude calls for the individual processes (core
process and each decay process) are generated separately, which allows
for very efficient treatment of multiprocesses, where multiple
processes have the same decaying final state. As a simple example consider
$pp\to W^+$ with $W^+\to l^+\nu_l$, which gives the core processes
$u\bar d\to W^+$, $\bar du\to W^+$, $c\bar s\to W^+$ and $\bar sc\to
W^+$, and the decay processes $W^+\to e^+\nu_e$, $W^+\to \mu^+\nu_\mu$
and $W^+\to \tau^+\nu_\tau$. The total number of subprocesses is obtained by 
combining  the four core processes with the three decay
processes. With the helicity amplitudes for each of the core processes
and each of the decay processes already generated, creating the
subprocesses only amounts to replacing the final state wavefunction
corresponding to the $W^+$ by the wavefunctions corresponding to the
decays into $e^+\nu_e$, $\mu^+\nu_\mu$ and $\tau^+\nu_\tau$
respectively. 

The procedure gets slightly more complicated when decay
processes have multiple diagrams- in this case, the diagrams for
the core process need to be multiplied together with the diagrams for the
decay.

The resulting matrix elements contain full spin correlations and
Breit-Wigner effects, but are not valid far from the mass peak, where
non-resonant diagrams might give significant interference effects. The tails of the
Breit-Wigner distributions for the specified decaying particles are
therefore cut off in \madevent, using the run parameter
\texttt{bwcutoff}, at $M\pm\Gamma*$\texttt{bwcutoff} (by default set to 15).

Process generation, as well as event generation with \madevent, has been
successfully tested up to 14 final state particles at the time of
writing this paper.

\section{Outputs}
\label{sec:outputs}

Previous versions of \madgraph\ could only output matrix
elements in Fortran~77. The modularized design of \madgraph~5
allows easy implementation of outputs in any language or user
desired format. The \ufo\ output and \aloha\ implementations allow for the
same flexibility in the output of models and helicity amplitude routines.

At present, matrix element output is available in Fortran~77, C++ and
Python. The Fortran~77 output is in the form of \madevent\ directory output (see
Sec.~\ref{sec:multiprocess}), or standalone matrix
element evaluation in the form of \madgraph\ standalone directory
output. For C++, presently available output formats are
standalone matrix element evaluation output, and dedicated output for \pythia~8 
(see Sec~\ref{sec:pythia8}). The Python matrix element code output is
currently used internally in \madgraph~5 to perform model
consistency checks during process generation as described in
Sec.~\ref{sec:model_checks} below. Implementation of other output formats, either in any of the currently
supported languages or different ones, can be easily done based on the 
existing output format implementations.

Finally, one of most useful outputs of \madgraph\ has always been the drawings
of the Feynman diagrams. Current new features (\eg , generation of processes with long decay chains)
as well as planned ones (\eg , generation of loop diagrams) required a new algorithm 
to be implemented. This is described in the last subsection.

\subsection{Multiprocess generation and MadEvent event generation}
\label{sec:multiprocess}

Simultaneous generation of multiple processes (e.g.\ $p p \to j j$,
using multiparticle
labels such as $p/j = g/d/u/s/c\bar/ d/\bar u/\bar s/\bar c$) has been
considerably optimized in \madgraph~5, using improved algorithms both for
process generation and for event generation.

On the process generation side, the time spent attempting to
generate processes which have no diagrams (such as $u \bar d \to g g$)
is minimized on one hand by checking any conserved quantum numbers, such as
electrical charge, provided by the model, and on the other hand in a
model-independent manner by keeping a memory of failed processes, and
ignoring any process which corresponds to a crossing of such a
failed process. This is
done before applying crossing symmetry breaking conditions such as required or forbidden
$s$-channel propagators. Furthermore, processes 
with mirrored initial state (such as $g u \to \gamma u$ and
$u g \to \gamma u$) are automatically recognized and combined into a
single process. To further speed up the generation, diagrams from
crossed processes are reused in the diagram generation (so that the
diagrams for, e.g., $g g \to u \bar u$, $u g \to u g$ and $u\bar u
\to g g$ are generated only once, and then reused with leg numbers
replaced as needed). 

The organization of subprocess directories for multiprocess event
generation in \madevent\ has 
been revamped. While \madevent\ 4 was already combining processes with
identical matrix elements (such as $g g \to u \bar u$ and $g g \to d
\bar d$), \madgraph~5 combines all processes with the same spin, colour and mass of
external particles into a single subprocess directory. The diagrams of these subprocesses
are matched to combined integration channels, so that a single
integration channel will perform single diagram enhanced phase space
integration \cite{Maltoni:2002qb} for all subprocesses with the corresponding
diagram. Diagrams whose pole structure differ only by permutations of
the final state momenta are also combined, to further minimize the
number of integration channels.  This means that for a process like $p
p \to l^+ l^- + 3j$, there will be only 5 subprocess directories,
corresponding to the subprocess groups $g g\to l^+ l^- g q q$, $g q\to
l^+ l^- g g q$, $g q\to l^+ l^- q q q$, $q q\to l^+ l^- g g g$, and $q
q\to l^+ l^- g q q$, as compared to 86 directories in \madgraph~4 (see
table~\ref{tab:grouping_comparison}).

\begin{table*}
\centering
\begin{tabular}{|c|c|c|c|c|c|c|c|c|}
\hline
\multirow{2}{*}{Process} & \multicolumn{2}{|c|}{Subproc. dirs.} &
\multicolumn{2}{|c|}{Channels} &
\multicolumn{2}{|c|}{Directory size} &
\multicolumn{2}{|c|}{Event gen. time} \\
\cline{2-9}
 & MG~4 & MG~5 & MG~4 & MG~5 & MG~4 & MG~5 & MG~4 & MG~5 \\
\hline
$pp\to W^+j$ & 6 & 2 & 12 & 4 & 79 MB & 35 MB & 3:15 min & 1:55 min\\
$pp\to W^+jj$ & 41 & 4 & 138 & 24 & 438 MB & 64 MB & 9:15 min & 4:19 min\\
$pp\to W^+jjj$ & 73 & 5 & 1164 & 120 & 842 MB & 110 MB & 21:41 min* & 8:14 min*\\
$pp\to W^+jjjj$ & 296 & 7 & 15029 & 609 & 3.8 GB & 352 MB  & 2:54 h* & 46:50 min*\\
$pp\to W^+jjjjj$ & - & 8 &- & 2976 & - & 1.5 GB & - & 11:39 h*\\
$pp\to l^+l^-j$ & 12 & 2 & 48 & 8 & 149 MB & 44 MB & 21:46 min & 3:00 min\\
$pp\to l^+l^-jj$ & 54 & 4 & 586 & 48 & 612 MB & 83 MB & 2:40 h & 11:52 min\\
$pp\to l^+l^-jjj$ & 86 & 5 & 5408 & 240 & 1.2 GB & 151 MB & 49:18 min* & 16:38 min* \\
$pp\to l^+l^-jjjj$ & 235 & 7 & 65472 & 1218& 5.3 GB & 662 MB & 7:16 h* & 2:45 h*\\
$pp\to t\bar t$ & 3 & 2 & 5 & 3 & 49 MB & 39 MB & 2:39 min & 1:55 min\\
$pp\to t\bar tj$ & 7 & 3 & 45 & 17 & 97 MB & 56 MB & 10:24 min & 3:52 min\\
$pp\to t\bar tjj$ & 22 & 5 & 417 & 103 & 274 MB & 98 MB & 1:50 h & 32:37 min\\
$pp\to t\bar tjjj$ & 34 & 6 & 3816 & 545 & 620 MB & 209 MB & 2:45 h* & 23:15 min*\\
\hline
\end{tabular}
\caption{\label{tab:grouping_comparison}
Number of subprocess directories, number of integration channels for the initial
run (``survey'') of the event generation, size of the
directory after one run generating 10,000 events, and run times for
generating 10,000 events, comparing
\mgme\~4 output (``MG~4'') with grouped subprocess output (``MG~5''). For
all processes, $p=j=g/u/\bar u/c/\bar
c/d/\bar d/s/\bar s$, $l^\pm=e^\pm/\mu^\pm$. The run times for 0-, 1- and 2-jet processes
are for a Sony VAIO TZ laptop with 1.06 GHz Intel Core Duo CPU running
Ubuntu 9.04, gFortran 4.3 and Python 2.6, while the 3-, 4- and 5-jet run
times (marked by *) are for a 128-core computer cluster with Intel Xeon
2.50GHz CPUs. $pp\to W^++5j$ is not possible to run with \mgme~4.
}
\end{table*}

Taken together with the identification of initial state mirror
processes, the number of integration channels for the initial cross
section determination run (``survey'') is significantly reduced, see
Table~\ref{tab:grouping_comparison}. Also for the subsequent event
generation run (``refine''), the number of integration channels is
usually reduced.  The required disk space is reduced by a factor
corresponding to the reduction in number of integration channels.

To further improve parallel running of the resulting configurations,
we have implemented the ability to split up channels with large
contribution to the cross section into multiple sub-channels, each generating a fraction of
the events for the channel in question. This results
in shorter generation times and more equal work load for jobs
submitted to a cluster, as well as considerably more stable
unweighting efficiency for the integration. Many further measures to
speed up and improve the stability of the generation have also been
taken, including a new initial guess for the shapes of the integration
variables, leading to considerably better stability of the
results. The resulting reduction in run times for a few sample
processes are also given in Table~\ref{tab:grouping_comparison}.

\subsection{Matrix element libraries for \pythia~8}
\label{sec:pythia8}

\pythia\ is one of the most widely used multipurpose event generators,
which includes matrix element evaluation, parton showering,
hadronization, particle decays and underlying events in a single framework. Matrix elements
for \pythia\ have historically been implemented by hand. The most
recent implementation of \pythia, the C++ version \pythia~8, allows
matrix elements for $2\to1$, $2\to2$ and $2\to3$ processes to be
provided by external programs. The flexibility in output formats 
in \madgraph~5 has allowed us to implement dedicated matrix element 
output for \pythia~8, thereby effectively removing the need for 
implementation of any matrix elements for \pythia\ by hand. 

Let us now describe the main features of this new implementation
for the readers interested in more technical details.

The new matrix elements are given in the form of classes inheriting from
the internal base class \texttt{SigmaProcess}. Such process classes
need to implement a number of member functions, providing \pythia\
with information about the process (initial states, external particle masses,
$s$-channel resonances, etc.), as well as functions to evaluate the
matrix elements for all included subprocesses and select final-state
particle id's and colour flow for each event. During event generation, 
\pythia\ 8 calls the matrix element classes with given momenta for 
the external particles. Starting from v.8.150,
\pythia\ also makes model parameters for new models (read in using the
Les Houches interface for BSM model parameters~\cite{Alwall:2007mw})
available to the resulting process class through an instance of the
class \texttt{SusyLesHouches}, allowing for the implementation and
simulation of processes in any new physics model.

The resulting matrix elements are treated by \pythia\ in exactly the same way as the
processes available in the \pythia\ core code. \madgraph~5 also
provides classes for evaluation of all model parameters needed for the
matrix element evaluation as well as the helicity amplitudes used by
the matrix elements. Standard model parameters are extracted from
internal \pythia\ parameters, while new physics parameters are read in
from BSM-LHA files. Just as in the regular \madevent\ output,
any model parameters based on the strong coupling constant
are recalculated event by event to use the running value of $\alpha_s$
(as provided by \pythia), while fixed model parameters are initialized
at the time of process initialization.

The \pythia~8 output of \madgraph~5 is a library called
\texttt{Processes\_\em{model\_name}}, which is automatically created
in a new directory under the \pythia~8 base directory. This library
contains source code files for all generated processes, model
parameters, and helicity amplitudes, as well as a makefile to compile
the library and place it in the \texttt{lib} directory of \pythia. As
an extra help to the inexperienced user, an example main program file
is also created in the \texttt{examples} directory together with a
dedicated makefile, which shows how to generate events from the implemented
processes. These main program files can be edited, compiled, and run
directly from the \madgraph~5 command line by running the
\texttt{launch} command, or compiled and run externally.

If multiple processes are generated in \madgraph~5, those processes
will automatically be arranged in different process classes according
to the initial and final states. This ordering uses the same machinery
as the new organization of \madevent\ subdirectories described in
Sec.~\ref{sec:multiprocess} above. In order for \pythia\ to perform
event generation, it needs all subprocesses inside a given process to
have the same spin and mass of external particles. 
This combination of subprocesses into single classes allows for
further optimization of the matrix element calculation, in that the
helicity wavefunctions and amplitudes can be calculated once and for all, and then be reused
between all the different subprocesses in a process class.

As a cross check, we have compared the cross section results for the
automatically generated processes with internal \pythia\ processes for
a large variety of processes in the Standard Model and the MSSM, with
perfect agreement.

\subsection{Diagram drawing}

In \madgraph~4 the amplitude generation and diagram drawing is limited to handling three and four point vertices. In addition,  the diagram drawing is based on a length minimization procedure, whose convergence is not a priori guaranteed and  sometimes creates lines with zero length. To overcome
these limitations, a completely new algorithm has been implemented in \madgraph~5. 
The basic idea is to associate to each vertex a level (\ie ,  to organise the verteces in classes each one characterized by 
a ``distance'' from the left end of the diagram) defined by the following simple rules:
\begin{itemize}
\item The vertices associate to initial particles are always at level one.
\item All vertices attached to a $t$-channel propagator are set at level one.
\item The difference of level between the endpoints of an $s$-channel propagator equals one.
\end{itemize}

As a boundary condition, all external lines are associated to a level at the start of the procedure: 
initial state particles are set to level zero, while all final state particle's 
levels are set to the maximal possible one plus 
unity.\footnote{Options are available that allow to modify the level of the external partons edges, resulting in different-looking graphs.}

The position of the vertices are then computed such that all vertices
at a given level lie equally spaced on the same vertical line. 

The main challenge of this method consists in avoiding line crossing
between propagators/final state particles.
In most situations, a simple re-ordering of the vertices at each level is enough to avoid line crossings.
The order should be such as vertices connected to the first vertex of
the previous level comes first, then the ones linked to the second one
and so on.
If a line connects two non-adjacent levels, which happens for final state particles, 
this methods needs an additional trick. Then, the intermediate level needs a special configuration 
-- \ie,  not equally spaced -- in order to avoid any possible line crossing. 
This is dealt with by adding a fake vertex (which is not drawn) to the intermediate level. 
Since we keep the ``equally spaced'' rule with the fake vertex, this creates the appropriate gap.
The resulting algorithm is very fast and efficiently generates clean diagrams with any number of external particles.

\section{Models}
\label{sec:models}

The \madgraph~5 library of models is built upon that of \feynrules\ \cite{Christensen:2008py}
and written in the  \ufo\ format \cite{ufo:2011}. Backward compatibility with the current
models of \madgraph~4 is supported as long as no particles with spin $3/2$ or higher are present in the model.

The set of currently publicly available models from the \feynrules\
wiki page is shown in the Table.~\ref{tab:models}. 
Several models are currently available in the \madgraph\ 5 release -
these are marked with a ``\checkmark'' in the table. Besides the
models in the table, also some additional models used for examples in this paper
are included in the release, e.g., the four-fermion interaction models
used in Sec.~\ref{sec:4fermion}.

\renewcommand{\tabcolsep}{1.5mm}
\TABLE{
\hspace*{-1cm}
\begin{tabular}{|c|c|c|c|}
\hline
Model Class & Model name &  Description/Comments & In MG5 \\
\hline
\multirow{2}{*}{Standard Model}
& sm & Several restrictions/simplifications 
available & \checkmark \\
\cline{2-4}
& heft & Top and W loops for the Higgs through dim-5 operators & \checkmark\\
\cline{2-4}
 \hline
\multirow{7}{*}{SM extensions}
 & 4Gen& Fourth Generation model with full CKM4 & \checkmark\\
 \cline{2-4}
 & SMScalars & Extra $O(n)$  scalar sector & \checkmark\\
\cline{2-4}
 & Hidden & Hidden Abelian Higgs Model & \\
\cline{2-4}
 & Hill & Hill Model &\\
 \cline{2-4}
  & 2HDM & The general Two Higgs-Doublet Model  & \checkmark \\
\cline{2-4}
  & TripletDiquarks & SM plus triplet diquarks & $\checkmark^*$\\
\cline{2-4}
  & SextetDiquarks &  SM plus sextet diquarks & $\checkmark^*$\\
 \hline
\multirow{3}{*}{SUSY models}
  & mssm & Minimal Supersymmetric extension of the SM  & \checkmark\\
\cline{2-4}
 & nmssm & Next-to-Minimal MSSM  & \checkmark\\
 \cline{2-4}
 & rmssm & $R$-symmetric MSSM & \\
 \cline{2-4}
 & rpvmssm & $R$ parity violating MSSM & \\
\cline{2-4}
\hline
\multirow{5}{*}{Extra-Dim models}
 & 3-site & Minimal Higgless Model (3-site Model)&  \\
\cline{2-4}
 & MUED & Minimal UED &\\
 \cline{2-4}
 & LED &  Large Extra Dimensions&  \\
\cline{2-4}
 & RS &  Randall-Sundrum  & \checkmark\\
\cline{2-4}
 & HEIDI & Compact HEIDI  & \\
\cline{2-4}
\hline
\multirow{3}{*}{EFT's}
 & ChiPT & Chiral perturbation theory  & \\
\cline{2-4}
 & SILH & Strongly Interacting Light Higgs &\\
\cline{2-4}
 & MWT & Technicolor  &\\
\cline{2-4}
\hline
\end{tabular}
\caption{\label{tab:models}
Selection of models that are currently available in \feynrules\ and can be used in \madgraph~5.
The last column indicate model which are present by default in the
current release of \madgraph~5. An asterisk (${}^*$) indicates that
the model in the \madgraph~5 library is a simplified version of the
complete model.}
}

\subsection{Inheriting models from \feynrules\ : \ufo\ and \aloha\ }
\label{sec:ufo_aloha}

Any local quantum field theory can be identified by:

\begin{itemize}
\item A set of particles and their quantum numbers  (spin, charges, etc.).
\item A set of parameters (masses, coupling constants, etc.).
\item A set of interactions among the different particles. 
\end{itemize}

The most efficient, reliable and compact way to encode that information
is by directly writing a Lagrangian with  (matter and interaction) fields
carrying the desired quantum numbers and by using well-known text-book rules 
to extract the Feynman vertices. This is what \feynrules\ does in a fully
automatic way. Once the vertices are obtained, the issue of passing
this information to a matrix element generator like \madgraph\ arises.
For example, for \madgraph~4,   \feynrules\  writes the output files exactly in the 
format needed by the code. This procedure, however, in addition to being
very heavy to maintain for \feynrules\ developers, has the drawback that
it suffers from the same intrinsic limitations of the \madgraph~4 model format itself.

The purpose of the Universal FeynRules Output (\ufo\ )~\cite{ufo:2011} is  to  overcome possible
limitations due to specific matrix element generators and translate {\em all} the information 
about a given particle physics model into a Python module that can 
easily be linked to any existing code. This output is complete and independent of the 
matrix element generator, allowing full flexibility and improvements. It 
saves the model information in an abstract  (generator-independent) way in terms 
of Python objects and classes, which in case of \madgraph~5 can be directly linked to the code. 

Once the information of a model is available in the \ufo, it can be used by \aloha\, 
to automatically write the \helas \ library corresponding to the corresponding Feynman rules.
\aloha , which is written in Python,  produces the complete set of routines (wave-functions and amplitudes) that are needed for the computation of Feynman diagrams at leading as well as at higher orders. The representation is language independent and outputs in Fortran, C++, Python are currently available. 

In so doing all the intrinsic limitations regarding the possibility of generating arbitrary new
physics model process of \madgraph~4 are overcome. As already explained above, \madgraph~4 is based
on the \helas\ library that encodes a limited set of Lorentz structures. While extensions are possible and 
have been done for several important cases (such as spin-2~\cite{Hagiwara:2008jb} and spin-3/2~\cite{Hagiwara:2010pi} particles), they entail a tedious work of writing and testing all new routines by hand.  
\aloha\ via the \ufo\  fully automates this procedure. In addition,
complicated interactions that feature non-factorizable  colour and
Lorentz structures,  such as those showing up in the counterterms and
$R_2$ vertices at NLO~\cite{Draggiotis:2009yb},  which cannot be handled by \madgraph~4, are now fully supported.

\subsection{Model restriction files}
\label{sec:model_restrictions}

In phenomenology applications, it is often convenient  to fix some parameters at some given values (\eg,
masses / CKM parameters / couplings, etc.).
Such restrictions might allow important gains both in terms of speed and also in 
size of the generated matrix element code.
\madgraph~5  allows to restrict a given UFO model 
based on a numerical evaluation of all couplings with parameters from
a BSM-LHA file \cite{Alwall:2007mw}
(note that analytical model restrictions can also be performed
directly inside \feynrules).

From a given BSM-LHA file (that we call a \emph{restriction file}),
\madgraph\ can evaluate the value of the internal 
parameters and of the couplings. The model is then modified according to the following rules:

\begin{itemize}
\item All vanishing parameters are removed from the model, and all
  parameters with value equal to unity are fixed to this value.
\item All parameters belonging to the same \texttt{LHA} block with identical values are represented by a single parameter.
\item All couplings with identical values are represented  by a single coupling.
\item All interactions linked to a vanishing coupling are removed from the list of interactions.
\end{itemize}
These steps allow \madgraph~5 to optimise the matrix element output,
and the output of multi-process generation (see Sec.~\ref{sec:multiprocess}).

In order to avoid the user setting the  parameters of the model in a way
which is inconsistent with the restricted model, we also modify the
param\_card associated with the model, removing all parameters that have been 
 fixed by the restriction. The default value for all remaining
parameters is set to that given in the restriction file. Note that
care is needed when the user designs the restriction file, to ensure
that the result corresponds to what it is expected, and that the resulting
param\_card includes all desidered parameters.

Model restrictions are used for several models present in
\madgraph~5, including SM and MSSM.
By default when a model is loaded, \madgraph~5 applies the restriction
defined in the file \texttt{restrict\_default.txt} in the
corresponding model directory.
For the Standard Model, the default restriction sets the CKM matrix to be diagonal and sets the mass of the first and second generation fermions to zero.
The user can bypass the default restriction by adding ``\texttt{-full}'' to
the model name when importing the model, or apply a different
restriction file by adding ``\texttt{\em -restriction}'' to the model
name, corresponding to a restriction file \texttt{restrict\_\emph{restriction}.dat}.

\subsection{Consistency checks for processes and models}
\label{sec:model_checks}

For new model implementations, either from \feynrules\
\cite{Christensen:2008py} or by directly providing the \ufo\  model format
used in \madgraph~5, it is crucial to be able to check the consistency
of the new models.
To this end, \madgraph~5 features a series of
consistency checks for processes:
\begin{enumerate}
\item Helicity amplitude calls and the helicity amplitude
implementations are checked by calculating the specified processes
with multiple permutations of the external particles in the diagram
generation in a given phase space point, checking that the value of
the matrix element is identical for the different permutations. 
This efficiently checks several aspects of the model
implementation: the relation between the Lorentz and colour structures, the implementation of the related 
helicity amplitudes with different wavefunction decomposition, and the effects
of fermion flow violation and charge conjugation.
\item Gauge invariance is checked by calculating the matrix element in
  a random phase space point with  the wavefunction of an
external massless vector boson replaced by its momentum (for processes with
external massless vector bosons). If gauge invariance is
satisfied, the resulting matrix element is zero (within numerical precision).
\item Invariance of the matrix element by Lorentz transformations is checked by comparing
the matrix element value before and after a series of Lorentz boosts.
\end{enumerate}
We have found that  altogether, these checks provide a 
powerful way to validate model implementations in \madgraph~5.

\section{Validation and speed benchmarks}

\subsection{Validation}

Once the  checks  based on symmetries, gauge invariance,  and Lorentz
invariance described  above are satisfied, one can perform the ``physics"
validation  by comparing results for the evaluation of the matrix
element in given points of the phase space and/or integrated cross
sections with other generators or analytical calculations. To validate
both \madgraph~5 and the models provided with it,
 we have extensively compared  squared matrix elements computed  point-by-point in the phase-space with those obtained in \madgraph~4.

Since \madgraph~5 supports different input model formats (the
\madgraph~4 format models and the new \ufo\ models) and is able to create
output in different languages (Fortran~77, C++), we have compared the \madgraph~4 results with  \madgraph~5 in the following three configurations:
\begin{itemize}
\item Using the \ufo\ model as input and choosing to export the matrix
  element in Fortran 77. (In this context the helicity amplitude routines are created by \aloha).
\item Using the \ufo\ model as input and choosing to export the matrix
  element in C++. (In this context the helicity amplitude routines are created by \aloha).
\item Using the \madgraph~4 model as input (therefore the only output available is Fortran 77 and we use the \helas\ package).
\end{itemize}

A summary of the checks performed is presented in Table
~\ref{tab:validation}. Those processes (more than three thousand in total) are
all in perfect agreement between all three configurations and the \madgraph~4 value.

\begin{table*}
\centering
\begin{tabular}{|c|l|l|c|}
\hline
model & process class & information& number of processes \\
\hline
SM & $A A \to A A$ & Only first generation for fermions& 249 \\
SM & $A A \to A A$ &  No first generation of fermions & 589 \\
SM & $B B \to B B B$ &  &86 \\
SM & $B B \to B F_{1,2} F_{1,2}$ &  & 46\\
SM & $F_{1,2} F_{1,2} \to B F_{1,2} F_{1,2}$ &  & 40\\
SM &  $F_{1,2} F_{1,2} \to B F_{1,2} F_{1,2}$ &  & 216\\
SM & $B B \to B B B B$ & & 55 \\
\hline
MSSM & $ P P \to \chi^0 \chi^0$ &  & 50\\
MSSM & $P P \to \chi^+ \chi^-/ \tilde g \tilde g$ &   & 26 \\
MSSM & $P P \to \tilde L \tilde L $ & & 55\\
MSSM & $P P \to \tilde Q_{1,2} \tilde Q_{1,2}$ & & 188\\
MSSM & $P P \to \tilde Q_3 \tilde Q_3$ & & 71\\
MSSM & $\tilde Q_{1,2} \tilde Q_{1,2} \to \tilde L \tilde L$ &  & 208 \\
MSSM & $\tilde Q_{1,2} \tilde Q_{1,2} \to \tilde Q_{1,2} \tilde Q_{1,2}$ & & 285 \\
MSSM & $ P P \to \tilde Q_{1,2}  \tilde Q_{1,2} V$ & only SM vector bosons included & 564 \\
MSSM & $ V V \to V  \chi^+ \chi^-$ &  & 71 \\
MSSM &  $ P P \to L^+ L^- \chi^0 \chi^0$  &  & 200 \\
MSSM & $V V \to V V \chi^+ \chi^-$ & & 177\\
\hline
HEFT  &  $B B \to B B$ & including the CP-odd Higgs & 62 \\
HEFT  &  $g g \to H+ ng$ & $n = 1,2,3,4$ & 4 \\
\hline
RS & $ AA \to AA$  &  Only first generation for fermions & 362 \\
RS & $ F_3 F_3 \to AA$  &  & 248 \\
RS & $ F_{1,3} F_{1,3} \to F_{1,3} F_{1,3} B $ & & 452\\
\hline

\end{tabular}
\caption{\label{tab:validation}
 Classes of processes used to compare the output of
  \madgraph~5 with \madgraph~4. The different letters designate classes of particles:
$A$ contains all the particles of the model;
$F_i$ contains  the $i^{th}$ generation of fermions of the model (No
indices means all generations allowed);
$L^\pm$ contains all the leptons of the model;
$V$ contains all the vector bosons of the model;
$B$ contains all the bosons of the model (\ie, the vector and the scalar particles);
$\chi^0$ contains all the neutralinos;
$\chi^\pm$ contains all the charginos;
$\tilde Q_i$ contains the $i^{th}$ generation of squarks;
$\tilde L$ contains the full set of sleptons.
}
\end{table*}

In addition to the squared matrix element tests, we have also
performed a systematic comparison at the cross-section level for $2
\to 2$ and $2\to 3$ both in the SM and in the MSSM. This comparison 
was done with the Feynrules web validation interface~\cite{FRweb:2011}
with respect to  \madgraph~4, \comphep, and \whizard.

\subsection{Speed benchmarks - process generation}
{
It can be interesting to compare the time required for the generation of
complete \madevent\ directories for different processes in \madgraph~4 and
\madgraph~5. Note that this is the time for the generation of the matrix
element output code and additional files needed for phase space
integration and event generation, not the time for evaluating the
matrix element values using the generated code. A time comparison for
evaluation the generated matrix element is given in the next section.

Table~\ref{tab:proc_generation_speed} shows the time needed for a
number of example processes, including multiprocesses,
high-multiplicity final state processes, and decay chain processes.
As can be seen from the table, the main gains in speed are in
complicated processes: processes with a large number of
subprocesses due to large multiplicities of multiparticle labels (such
as $pp\to jjje^+e^-$), processes with many external particles (such as 
$gg\to5g$ or $e^+e^-\to6e$), and in decay chain processes, where
the speedup can be several orders of magnitude with respect to previous
versions of \madgraph.

\begin{table*}
\centering
\begin{tabular}{|c|c|c|c|c|}
\hline
Process & \madgraph~4 & \madgraph~5 & Subprocesses & Diagrams \\
\hline
$pp\to jjj$ & 29.0 s & 25.8 s & 34 & 307 \\
$pp\to jj l^+l^-$ & 341 s & 103 s & 108 & 1216 \\
$pp\to jjj e^+e^-$ & 1150 s & 134 s & 141 & 9012 \\
$u\bar u\to e^+e^-e^+e^-e^+e^-$ & 772 s & 242 s & 1 & 3474 \\
$g g\to g g g g g$ & 2788 s & 1050 s & 1  & 7245 \\
$pp\to jj(W^+\to l^+\nu_l)$ & 146 s & 25.7 s & 82 & 304 \\
$pp\to t\bar t+$full decays & 5640 s & 15.7 s & 27 & 45 \\
$pp\to\tilde q/\tilde g\,\tilde q/\tilde g$ & 222 s & 107 s & 313 & 475 \\
7 particle decay chain & 383 s & 13.9 s & 1 & 6 \\
$gg\to(\tilde g\to u\bar u\tilde\chi_1^0)(\tilde g\to u\bar u\tilde\chi_1^0)$ &
70 s & 13.9 s & 1 & 48 \\
$pp\to(\tilde g\to jj\tilde\chi_1^0)(\tilde g\to jj\tilde\chi_1^0)$ &
--- & 251 s & 144 & 11008 \\
\hline
\end{tabular}
\caption{\label{tab:proc_generation_speed}
Time for generation of complete \madevent\ directories (with
the exception of $gg\to5g$, for which a Fortran standalone directory was
generated) for a selection of processes, for
\madgraph~4 and \madgraph~5. All processes have $p=j=g/u/\bar u/c/\bar
c/d/\bar d/s/\bar s$, $l^\pm=e^\pm/\mu^\pm/\tau^\pm$,
$\nu_l=\nu_e/\nu_\mu/\nu_\tau$ and
$\bar\nu_l=\bar\nu_e/\bar\nu_\mu/\bar\nu_\tau$. $\tilde q/\tilde g$ in
the table corresponds to $\tilde d_{l/r}^{(*)}/\tilde
u_{l/r}^{(*)}/\tilde s_{l/r}^{(*)}/\tilde c_{l/r}^{(*)}/\tilde g$. For
$t\bar t+$full decays (meaning $pp\to (t\to b\,q/l^+\,\bar
q/\nu_l)(\bar t\to \bar b\,q/l^-\,\bar q/\bar \nu_l)$), the \madgraph~4
process generation was split up in 12 different process
definitions to reduce the number of failed process attempts. The
``seven particle decay chain'' was $gg\to(\tilde g\to u(\bar{\tilde
u}_l\to\bar u(\tilde\chi^0_2\to Z\tilde\chi^0_1)))(\tilde g\to u\tilde
d\tilde\chi_1^-)$. The number of subprocesses and diagrams are quoted after
combination of subprocesses with identical matrix elements. All
processes are generated with maximal number of QCD vertices.
 All numbers are for a Sony VAIO TZ laptop with
1.06 GHz Intel Core Duo CPU running Ubuntu 9.04, gFortran 4.3 and
Python 2.6.}  
\end{table*}

\subsection{Speed benchmarks - matrix element evaluation}
\label{sec:evaluation_speed}

\madgraph~5 is not only faster than its predecessors in generating code for
complicated processes, the produced matrix element code is also faster
and more compact. This is thanks to the new diagram generation
algorithm, which allows for improved recycling of subdiagram wavefunctions
between different diagrams, reducing the number of helicity
wavefunction calls (as discussed in Sections \ref{sec:diagram_generation} and
\ref{sec:fermionflows}). Table~\ref{tab:me_evaluation_speed} shows a
comparison of the number of function calls and run time
for matrix element evaluation using \helas\ and \aloha\ routines, relative to \madgraph~4.

\begin{table*}
\centering
\begin{tabular}{|c|c|c|c|c|c|}
\hline
\multirow{2}{*}{Process} & \multicolumn{2}{|c|}{Function calls} &
\multicolumn{2}{|c|}{Run time relative to MG~4} \\
\cline{2-3} \cline{4-5}
 &  MG~4 & MG~5 & MG~5+\helas &MG~5+\aloha \\
\hline
$u\bar u\to e^+e^-$ & 8 & 8 & 1.0 & 1.1 \\
$u\bar u\to e^+e^-e^+e^-$ & 110 & 80 & 0.52 & 1.4 \\ 
$u\bar u\to e^+e^-e^+e^-e^+e^-$ & 6668 & 3775 & 0.33 & 0.57 \\
$g g\to g g$ & 13 & 13 & 0.90 & 0.81 \\
$g g\to g g g$ & 86 & 78& 0.94 & 0.94 \\
$g g\to g g g g$ & 811 & 621 & 0.99 & 0.66 \\
$u\bar u\to d\bar d$ & 6 & 6 & 1.0 & 1.0 \\
$u\bar u\to d\bar d g$ & 16 & 16 & 1.0 & 1.2 \\
$u\bar u\to d\bar d g g$ & 85 & 67 & 0.74 & 0.86 \\
$u\bar u\to d\bar d g g g$ & 748& 515  & 0.68 & 0.52 \\
$u\bar u\to u\bar u g g$ & 160 & 116 & 0.67 & 0.70 \\
$u\bar u\to u\bar u g g g$ & 1468 & 960 & 0.48 & 0.36 \\
$u\bar u\to d\bar d d\bar d$ & 42 & 33 & 0.99 & 1.2 \\
$u\bar u\to d\bar d d\bar d g$ & 310 & 197 & 0.61 & 0.74 \\
$u\bar u\to d\bar d d\bar d g g$ & 3372 & 1876 & 0.24 & 0.19 \\
$u\bar u\to d\bar d d\bar d d\bar d$ & 1370 & 753 & 0.18 & 0.19 \\
\hline
\end{tabular}
\caption{\label{tab:me_evaluation_speed}
Number of helicity function calls and run time ratio to \madgraph~4 for
matrix element evaluation of matrix element code produced by
\madgraph~4 and \madgraph~5 using \helas, and \madgraph~5 using
\aloha\ routines. For \madgraph~4, the \helas\ library has been used. 
Note that in the $u\bar u\to q\bar q + X$ process generations, 
only QCD interactions have been allowed
(\texttt{QED=0}). The number of function calls for \madgraph~5 does
not depend on whether \helas\ or \aloha\ is used.
}  
\end{table*}

From the table, we see how the improved wavefunction call
optimisation translates to considerably improved run times, especially for complicated processes.

\section{BSM example applications}
\label{sec:examples}

\subsection{Non-standard colour structures: $\epsilon^{ijk}$ and colour
  sextets}

The phenomenology of diquark resonances has recently
become popular, since these particles could be among the first new
physics particles to be observed at the proton on proton collider LHC
\cite{Han:2009ya}. Such diquarks have the quantum numbers of two
valence quarks, and must therefore be either colour sextets or colour
anti-triplets. In the latter case, the coupling to quarks is through a
completely antisymmetric colour triplet $\epsilon$ tensor, which is the
only way to contract three colour triplet indices. The triplet
$\epsilon$ tensor is also important in the formulation of $R$-parity
violation supersymmetric models, where e.g.\ a scalar quark can decay
into a pair of Standard Model antiquarks. As discussed in
Sec.~\ref{sec:colour_algebra}, both the sextet colour algebra and the $\epsilon$
tensor have been implemented in \madgraph~5.

\begin{figure}
\centering
\includegraphics[width=6cm,angle=-90]{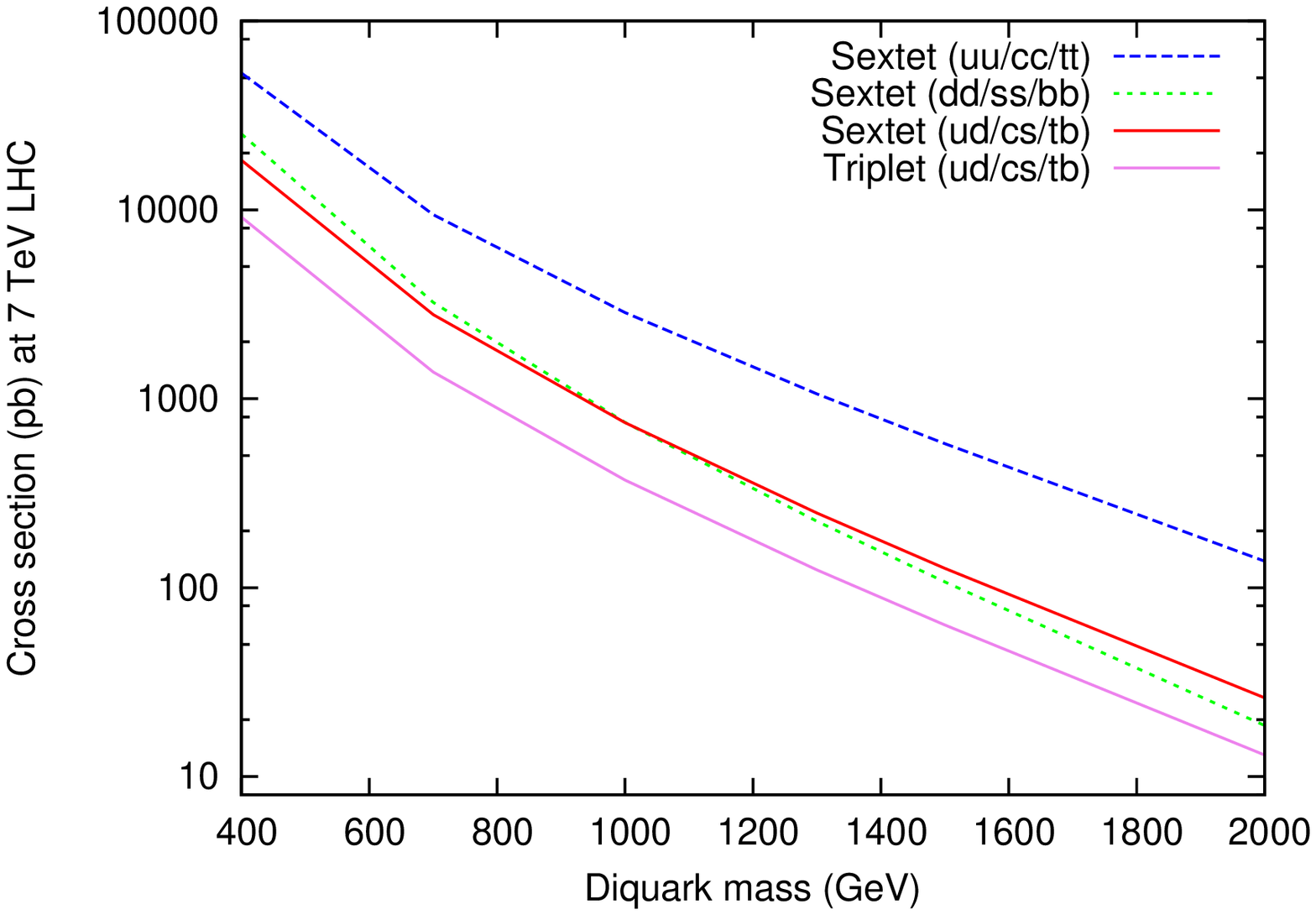}
\includegraphics[width=8cm]{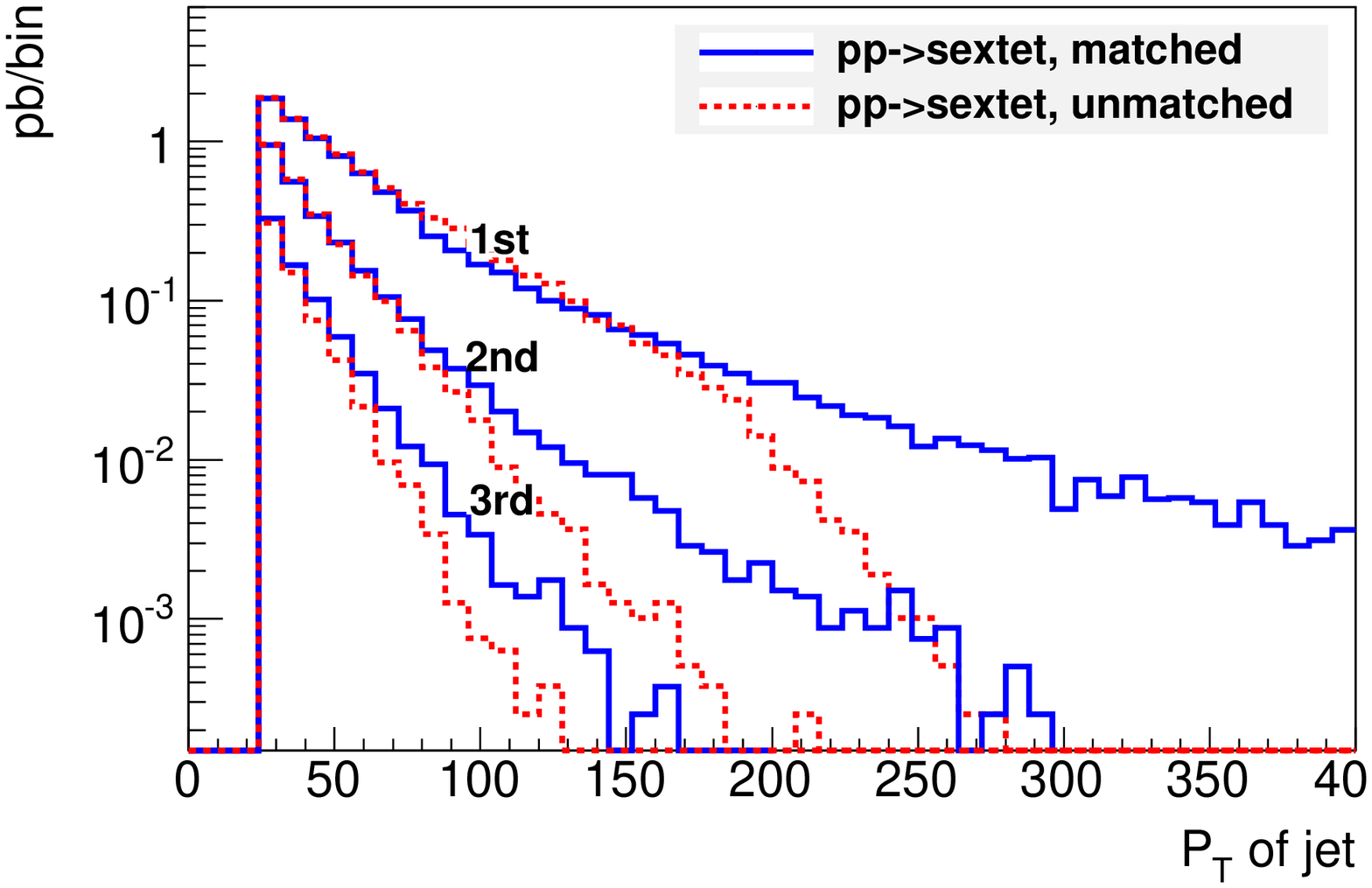}
\caption{Upper: Cross sections for different types of diquark
  resonances at 7 TeV LHC. See text for details. Lower: Comparison of
  $p_T$ for radiated jets
  between single colour sextet diquark production with jet matching
  (using \madgraph\ and \pythia) and without jet matching (using
  \pythia\ parton showers for the leading order process $p p\to D$
  only). See text for details.}
\label{fig:ptjets-comparison}
\end{figure}

As an example of phenomenology using these implementations, we show in
Fig.~\ref{fig:ptjets-comparison}a the cross sections for
different species of colour sextet and antitriplet scalar diquarks $D$ at LHC with 7 TeV
c.m.\ energy. 
We
have included colour sextet diquarks coupling to $uu$/$cc$/$tt$,
$dd$/$ss$/$bb$ and $ud$/$cs$/$tb$, and colour antitriplet diquarks
coupling to $ud$/$cs$/ $tb$. Note that due to the antisymmetry of the
$\epsilon^{ijk}$ colour coupling of colour triplet diquarks to quarks,
colour triplet diquarks can only couple to off-diagonal flavour quark
combinations. The $Dqq^{(\prime)}$ Yukawa coupling constants have been set to $10^{-2}$ in
the figure. Note the factor 2 between the $pp\to D$ production cross
sections for sextet and triplet diquarks (for identical Yukawa
couplings), due to the different colour factors. 

In Fig.~\ref{fig:ptjets-comparison}b, we show the effect of jet
matching between matrix elements and parton showers for charge
$+\frac43$ sextet diquark production at 7 TeV LHC. $p_T$ distributions
for the radiated jets are compared between matched production with
\madgraph\ (using the $k_T$-MLM matching scheme that is default in
MadGraph, with matrix elements for $p p\to D+0,1,2$ jets) and
\pythia\ 6.4 with $p_T$-ordered showers, and just using the leading
order process $p p \to D$ with $p_T$-ordered \pythia\ default
settings. The mass of the diquark is 500 GeV. It is clear that
matching is necessary for a precise description of high-$p_T$ jet
radiation in association with diquark production.

\subsection{4-fermion vertices: $u u \to t t$}
\label{sec:4fermion}

With the possibility of specifying vertices with arbitrary number of
particles, a particular difficulty arises when a vertex has more than
two fermions, in which case it is necessary to define the
fermion flow in an unambiguous way. The convention chosen by the
\madgraph~5 and \feynrules\ authors is that the fermion flow is
defined by the position of the particle in the interaction, with the
order being $IOIO\ldots$ where $I$ stands for ``incoming'' and $O$
stands for ``outgoing'' fermion. Any number of bosons can be added
after the fermions.

This means that, from the \madgraph~5 point of view, the interactions
$u \bar t u \bar t$ and $u u \bar t\bar t$ are treated in different
ways - in the former case, the fermion flows go between $u$ and $\bar
t$, while in the latter case we have fermion number violating flows
$u\to u$ and $\bar t\to \bar t$. Such fermion number violating
multi-fermion vertices are readily treated by the algorithm
described in Sec.~\ref{sec:fermionflows}, by the use of conjugate
$\Gamma$ matrices for each fermion number violating fermion line.

Of particular interest are four-fermion vertices, which are a common
feature of effective theory formulations for physics beyond the
Standard Model, and have recently been studied in the context of top-quark LHC
phenomenology \cite{Zhang:2010dr,AguilarSaavedra:2010zi,Degrande:2010kt}
We are here presenting two examples of four-fermion vertices leading to 
the process $u u \to t t$\cite{Degrande:2011rt}, one which corresponds to the exchange of a
heavy $s$-channel propagator (in this case a scalar colour sextet
diquark) and one which corresponds to a heavy $t$-channel propagator (a
neutral colour singlet flavour changing scalar). In both cases,
we use a mass of 10 TeV for the propagators, and represent the
four-fermion vertices by integrating out the appropriate scalar propagator.

In Fig.~\ref{fig:4fermion_comparisons} we compare the full theories,
including the explicit propagators, with the 4-fermion vertex versions
of the theories. The figure shows $u u \to t t$ cross sections for the
two scenarios as a function of fixed center of mass energy, for some
particular coupling values; to compare with a particular collider, the
cross sections need to be convoluted with the $uu$ parton luminosity.
The sudden turn-on of the cross section is due to the kinematical
suppression from the final-state top quark mass.  

As expected, the four fermion vertex formulation agrees with the
explicit propagator formulation up to a c.m. energy of about $1/10$ of
the propagator mass. The explicit $t$-channel propagator makes the cross
section level off as the energy gets close to the mass and the
exchange momentum term in the denominator of the propagator starts
dominating over the mass term, while the explicit $s$-channel propagator
displays the usual Breit-Wigner peak as the energy gets
close to the mass. In this case, the width of $s$-channel propagator is
$\Gamma_S = 200$ GeV.

\begin{figure}
\centering
\includegraphics[angle=-90,width=7cm]{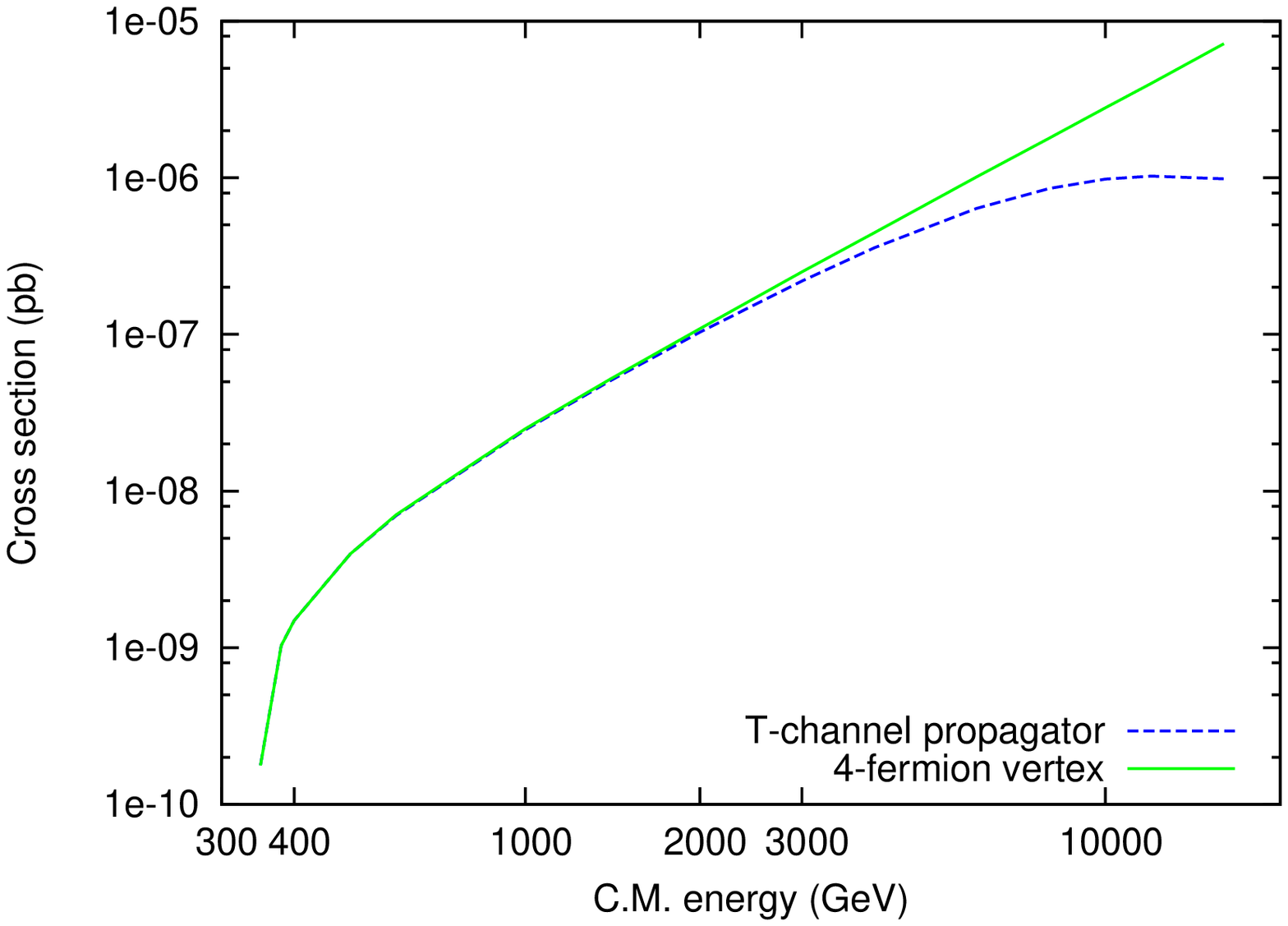}
\includegraphics[angle=-90,width=7cm]{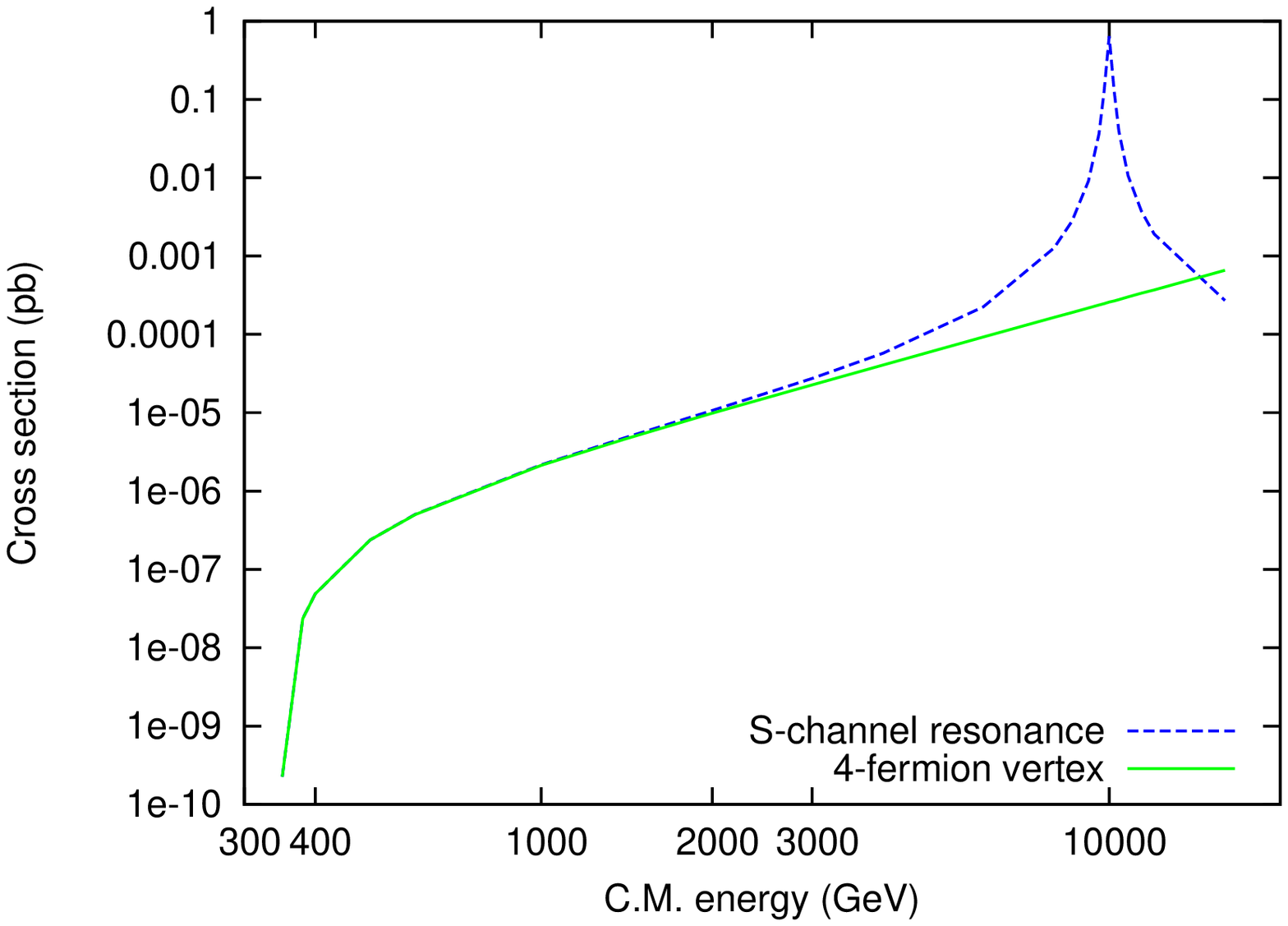}
\caption{Cross section for $u u \to t t$ as a function of (fixed) beam energy, comparing
  4-fermion implementations and the corresponding implementations with
  explicit propagators. Left: $t$-channel scalar exchange. Right:
  $s$-channel scalar exchange. For both cases, the mass of the
  propagator is set to 10 TeV.}
\label{fig:4fermion_comparisons}
\end{figure}

\subsection{$n$-particle vertices: $H+4g$}

\madgraph~5 allows vertices with any number of external
particles. Such vertices frequently appear in effective field
theories, where non-renormalizable operators are included with some
appropriate scale suppression. One of the most phenomenologically
important effective field theories for high-energy physics is the
addition to the Standard Model of effective couplings between the
Higgs boson and gluons through a top quark loop, with the top mass
taken much larger than the Higgs boson mass.

While the simplest vertex we can write down in this theory is $Hgg$,
the non-Abelian nature of QCD require us to include two additional
operators in the Lagrangian, $Hggg$ and $Hgggg$, coupling the Higgs
directly to
three and four gluons respectively. While previous versions of
\madgraph\ had to split up the five-particle $Hgggg$ vertex using an
auxiliary non-propagating tensor particle, \madgraph~5, in conjunction
with \aloha, can directly handle this vertex in exactly the same way
that it handles  vertices with lower multiplicity. The corresponding
diagram, from the process $gg\to Hgg$, is found in
Fig.~\ref{fig:gg_Hgg_diagram}.

Note that for consistency, only one effective Higgs-gluon coupling
vertex can be present in a given diagram. This is made possible by
specifying a separate coupling order, \texttt{HIG}, for these
vertices. Any process generation in this model should therefore be
specified with a maximum order \texttt{HIG=1}.

\begin{figure}
\centering
\includegraphics[trim=70 570 160 -280,width=6cm,clip=true]{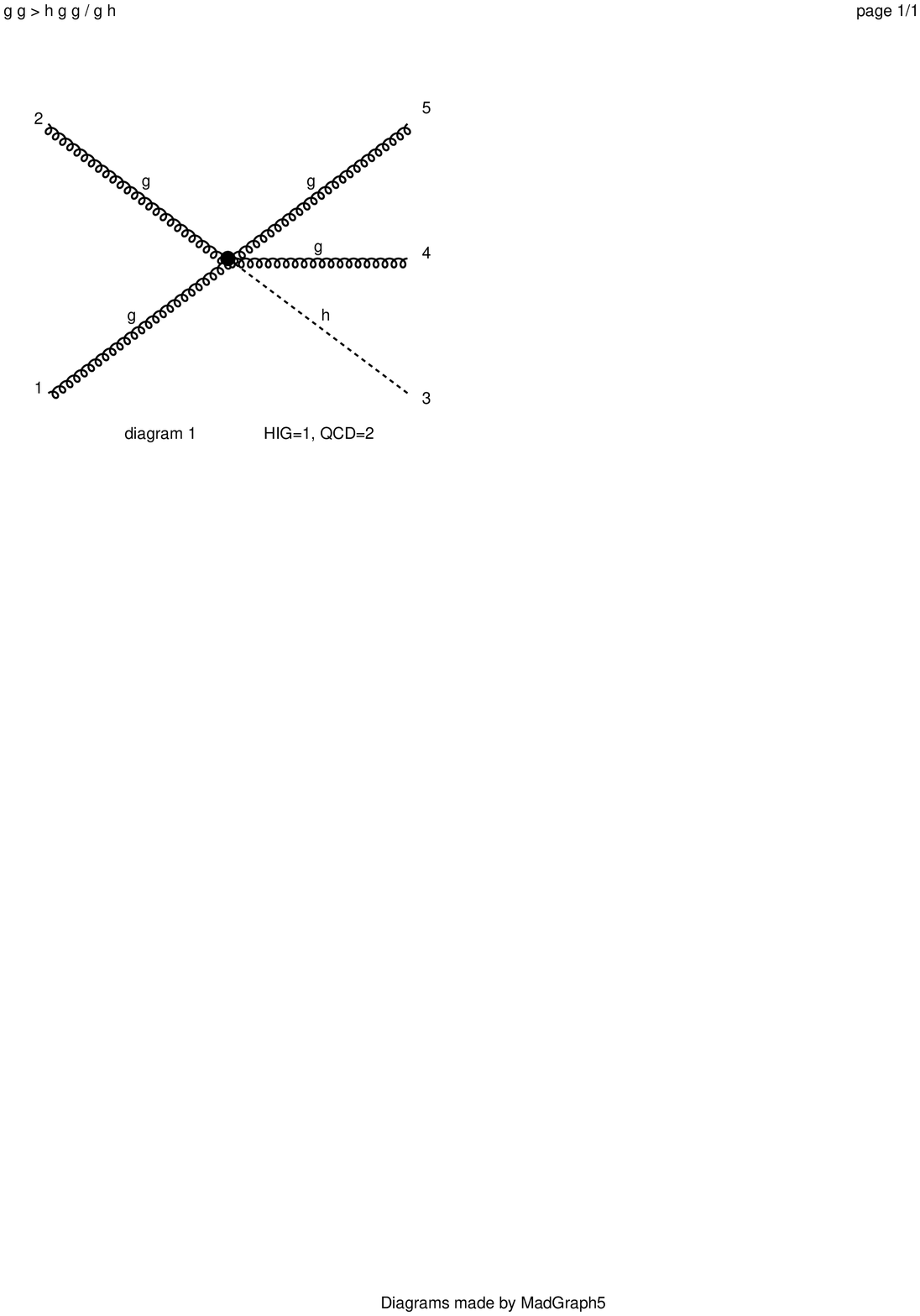}
\caption{5-particle diagram by \madgraph~5. This is one of 38 diagrams
  from the process $g g \to H g g$ in the Standard Model with
  effective couplings of the Higgs boson to gluons through a top loop.}
\label{fig:gg_Hgg_diagram}
\end{figure}

The new formulation of Higgs boson couplings to gluons has been
thoroughly checked against the implementation in \madgraph~4, and is
one of the models included in the \madgraph~5 package.

The same issue happens for a graviton interacting with four gauge bosons. 
A full validation of the  \madgraph~5  implementation the \feynrules\  RS model against the 
 \madgraph~4 \cite{Hagiwara:2008jb} implementation has been performed.

\subsection{Chromo-magnetic operator}

Recent measurements in top quark pair production performed both at Tevatron and LHC 
offer an ideal ground to search for new physics effects \cite{Degrande:2010kt}. If such new physics is at scales
higher than those explored at the current colliders it can be efficiently modeled by an effective 
theory with a non-renormalizable operator suppressed by a energy-scale $\Lambda$. 
In the case of the top quark, only one operator of dimension 6 exists that is not a 4-fermion operator, the so called 
\emph{chromo-magnetic}  operator:
\begin{equation}
\mathcal{L} = \frac{(H\bar Q)\sigma^{\mu\nu}T^At G^A_{\mu\nu}}{\Lambda^2} + h.c., \nonumber
\end{equation}
where $\Lambda$ represents the cutoff of the effective theory.
Adding such a term to the Lagrangian of the standard model leads to additional interactions, see Fig. \ref{fig:chromo_diag}. For consistency, any matrix element should  be computed up to order $\Lambda^{-2}$ and therefore requiring one and only one effective coupling to enter at the squared matrix element level. 
This can be obtained by defining a third type of coupling (in addition to QED and QCD) associated with the 
new interactions.

\begin{figure}
\centering
\includegraphics[trim=150 300 -40 -300,width=8cm]{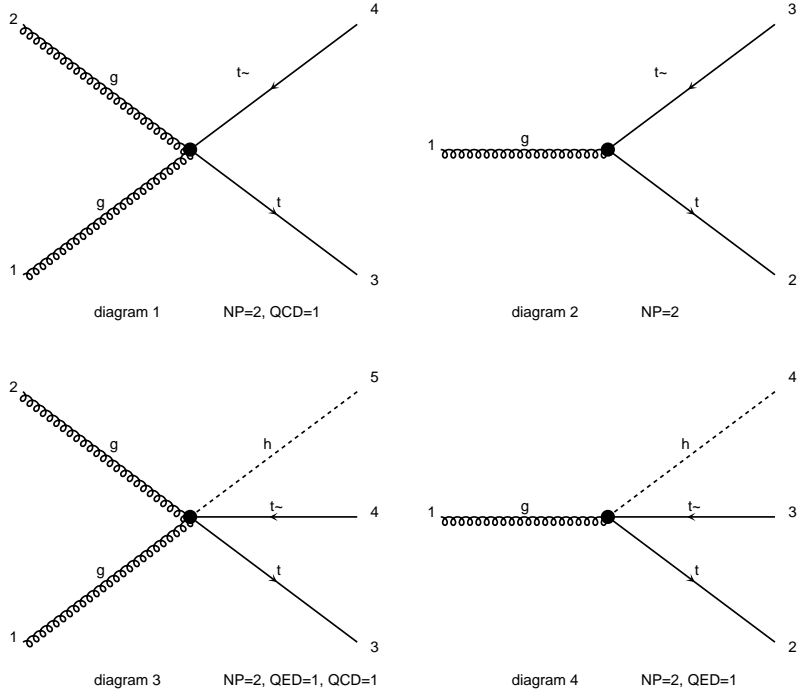}
\caption{Interactions induced by the chromomagnetic operator. }
\label{fig:chromo_diag}
\end{figure}

Again the automation of the \helas~routines and the possibility for
\madgraph~5 to deal with vertices with arbitrary number of legs,
allows a straightforward treatment of such operators. The validation
of this model has been performed by comparing the \madgraph~5 results
to those obtained in a private (and not fully automatic)
implementation of this model inside the \madgraph~4 framework. In
addition to automatically providing the matrix element for any process
involving the new operators, the
evaluation of the cross sections are approximatively four times faster than the previous version.

\section{Conclusions and Outlook}

The complete rewriting of the \madgraph\ matrix-element generator in Python has
allowed us to build on the twenty years of experience gained with the Fortran version
and push the code as well as its functionalities to a new level. The resulting code,
modular in structure and with embedded robustness and sanity checks, is naturally organised
as a collaborative platform. Any sufficiently skilled user can exploit, modify and 
extend the functionalities of the current version. The code is
available via a major open-source project development hosting service
using the Bazaar version control system.

The new code structure and new functionalities open the way to developments 
in three main directions: BSM, NLO, and merging with shower/hadronization codes.
The direct link to the \feynrules\ model database 
will allow  quick and robust implementation not only of new physics models but also
of any type of 1-loop counterterms, essential ingredients to achieve NLO automatic computations
in the SM and beyond.  Work in all these directions is in progress.

\section*{Acknowledgements}

It is a great pleasure for us to thank all the people who, directly or indirectly, help and support our efforts and services to the high-energy community and all our users for their continuous and patient feedback.  In particular, for the help in the extensive testing of the new version, we thank Alexis Kalogeropoulos; for the validation of new physics models (and much more) we thank the \feynrules\ core authors (Neil Christensen, Claude Duhr, Benjamin Fuks) and associates (Priscila de Aquino, Celine Degrande); for our cluster management we are grateful to Vincent Boucher, Jer\^ome de Favereau, Pavel Demin, and Larry Nelson; for the great physics work and the fun, we in particular thank our colleagues and collaborators: Pierre Artoisenet, Simon de Visscher, Rikkert Frederix, Stefano Frixione, Nicolas Greiner, Kaoru Hagiwara, Junichi Kanzaki, Valentin Hirschi, Qiang Li, Kentarou Mawatari, Roberto Pittau, Tilman Plehn,  Marco Zaro. This work is partially supported by the Belgian Federal Office for Scientific, Technical and Cultural Affairs through the 'Interuniversity Attraction Pole Program - Belgium Science Policy' P6/11-P and by the IISN ``MadGraph" convention 4.4511.10.

\appendix

\section{Installation and Online Web Version}

\madgraph~5 can be either used directly online on the web or locally. 
The code can be downloaded from the web page {\tt https://launchpad.net/madgraph5}. 
In order to run \madgraph~5 locally, \python~2.6 or higher (but not 3.x) must be installed.
The package does not require any compilation or configuration; after
unpacking, simply launch the main script:
\begin{verbatim}
tar -xzpvf MadGraph5_v1.x.x.tar.gz
cd MadGraph5_v1_x_x
./bin/mg5
\end{verbatim}
To learn how to use \madgraph~5, enter \texttt{tutorial} in the
command line interface.
The full list of presently available commands are described in App.~\ref{app:command}.

\madgraph~5 can also be used on the web (after free registration). We currently have three public clusters:
\begin{itemize}
\item {\tt http://madgraph.phys.ucl.ac.be/}
\item {\tt http://madgraph.hep.uiuc.edu/}
\item {\tt http://madgraph.roma2.infn.it/}
\end{itemize}
At this stage, \madevent\ (Fortran) output can be generated online, and
downloaded as a standalone process directory. In the near future, also
other output format will be available online.
Based on a user request, we also grant access to run event generation
directly on one of the clusters.
In that case, the user can also choose to directly pass the events
through \pythia~6 for hadronization, and use a fast detector simulation, either \delphes~\cite{Ovyn:2009tx} or \pgs~\cite{PGS}.

\section{Command line user interface}
\label{app:command}

The command line interface for \madgraph~5 is built on the Python module
\texttt{cmd}. This module allows for a flexible treatment of user
input and support of features such as tab completion, command
history (accessed by the \texttt{up} key), help texts, and access to
shell commands from inside the command line interface. Using the
command line interface, the user can conveniently access the full functionality of 
\madgraph~5, including importing models, generating processes, drawing
Feynman diagrams,
generating output in all available output formats (at present,
\madevent, \pythia~8 and standalone process and model output
in Fortran or C++), performing model checks, and launching event
generation in previously created process directories.

The command line interface is straightforward to extend, and more
functionality is continuously added based on user requests and code development.

The syntax for process generation in \madgraph~5 is very similar to
that for \madgraph~4, with a few exceptions reflecting the extended
functionality of \madgraph~5 -- most notably, spaces are needed
between particle names, since there is no longer any limit on particle
name length, and the syntax for generating decay chains is modified to
accommodate the greater flexibility in decay chain generation. Some
syntax examples are given in table~\ref{tab:syntax_examples}. However, 
the interface also has the ability to read process cards written for \madgraph~4.

\begin{table*}
\centering
\begin{tabular}{|p{5cm}|p{9cm}|}
\hline
Syntax & Description \\
\hline
\texttt{p p > l+ l-} & Generate the (valid) processes $u\bar u \to e^+e^-$,
$d\bar d \to \mu^+\mu^-$ etc.\\
\hline
\texttt{p p > j j j} & (No specification of orders.) Only
allow diagram with maximal \texttt{QCD}/minimal \texttt{QED} order.\\
\hline
\texttt{p p > j j j QED=2} & Allow up to 2 \texttt{QED} vertices (and
unlimited QCD vertices) in
diagrams.\\
\hline
\texttt{p p > z > l+ l- j} & Include only diagrams with an $s$-channel $Z$.\\
\hline
\texttt{l+ l- > z | a > l+ l-} & Include only diagrams with an
$s$-channel $Z$ or an $s$-channel $\gamma$.\\
\hline
\texttt{b b\urltilde\ > t t\urltilde\ / g} & Exclude any diagrams with a $g$ as
internal propagator.\\
\hline
\texttt{p p > b w+ t\urltilde\ \$ t} & Exclude any diagrams with an $s$-channel $t$ propagator.\\
\hline
\begin{minipage}[t]{1.0\columnwidth}
\texttt{p p > t t\urltilde, t > b w+, \textbackslash} \\ 
\texttt{(t\urltilde\ > b\urltilde\ w-, w-
  > l- vl\urltilde)}   
\end{minipage}
& Generate a decay chain with $t$, $\bar t$ and $W^-$ required to be near-onshell. Note the use of
parentheses to specify decays within a decay chain.\\
\hline
\end{tabular}
\caption{\label{tab:syntax_examples} Some examples of process
  generation syntax in the \madgraph~5 command line interface. The
  main differences w.r.t.\ \madgraph~4 are that spaces are needed
  between particle names, that by default minimal \texttt{QED}
  coupling order is assumed (if there are only \texttt{QED} and
  \texttt{QCD} couplings in the model), and furthermore the decay
  chain syntax which now allows full specification of all decay
  processes including coupling orders, required and excluded
  particles, etc.}  
\end{table*}

The user can start the command line interface by running \texttt{bin/mg5} from the
\madgraph~5 directory. If the name of a file containing \madgraph~5
commands is given as argument, then the commands in the file are
performed. Such a file can be generated from the series of commands
used in a session by the \texttt{history} command, and is automatically
placed in the \texttt{Cards/} directory when a \madevent\ directory is
created. Process generation can also be done as in \madgraph~4, by
running \texttt{bin/newprocess\_mg5} in a copy of the Template
directory with an appropriate \texttt{proc\_card.dat} or \texttt{proc\_card\_mg5.dat} placed in
the \texttt{Cards} directory.

We list here the presently available commands in the command interface
(in alphabetical order). 

\begin{itemize}
\item \texttt{add process}: Add and generate diagrams for a process,
  keeping previously generated processes.
\item \texttt{check}: Run model consistency checks for
  specified processes. Available checks are: process permutation
  checks, gauge invariance check, and Lorentz invariance check (see
  Sec.~\ref{sec:model_checks}).
\item \texttt{define}: Define a multiparticle label used for implicit
  summing over processes. Some commonly used multiparticle labels
  (\texttt{p}, \texttt{j}, \texttt{l+}, \texttt{l-}, \texttt{vl},
  \texttt{vl\urltilde}) are automatically defined when a model is imported.
\item \texttt{display}: Display particles, interactions, defined
  multiparticle labels, generated processes, generated diagrams, or
  the results of process checks.
\item \texttt{generate}: Generate diagrams for a process, replacing any
  previously generated processes.
\item \texttt{history}: List the history of previous commands to the
  screen or to a file. The resulting file can be used to repeat a
  sequence of commands using the \texttt{import} command.
\item \texttt{import}: Import a model (either in the \ufo\ format
  or \madgraph~4 format) or a process card (in \madgraph~5 or
  \madgraph~4 format).
\item \texttt{launch}: Launch event generation or matrix element
  evaluation for created process directories.
\item \texttt{load}: Load a process or model previously saved to file
  using the \texttt{save} command.
\item \texttt{output}: Output process files for matrix element
  integration. Presently available output formats are: \madevent\
  format (default), stand-alone Fortran format, \pythia~8 format, and
  stand-alone C++ format.
\item \texttt{save}: Save a process or model to file, using the Python
  ``pickle'' format.
\item \texttt{set}: Modify settings, including turning on/off
  subprocess grouping (see Sec.~\ref{sec:multiprocess}).
\item \texttt{tutorial}: Start a short tutorial showing how to use the
  most common commands.
\end{itemize}

Note that the \texttt{help} command gives the full list of available
commands. Typing \texttt{help \emph{command}} gives information about
each command. The built-in commands \texttt{quit} and
\texttt{exit} (or pressing \texttt{control-D}) quits \madgraph~5, and
\texttt{shell} (or starting the line with a ``\texttt{!}'') allows to
access any shell command.

Finally, the interactive interface has a tutorial mode which allows to
quickly learn the most common commands used in the interface.

\section{Process generation examples}

This appendix presents examples of the series of commands required to generate the code corresponding to 
the square matrix element of various process.  Those command can be either copied-pasted directly in 
the interactive interface (accessed by running \texttt{./bin/mg5}) or
written in a text file executed by \madgraph~5 (executed by \texttt{./bin/mg5 \emph{command\_file}}).

\subsection{Top-quark pair production}
\label{subsec:manual_tt}

The first example shows how to evaluate a cross-section (and how to generate partonic events) for top-quark pair production in the Standard Model:
\begin{verbatim}
generate p p > t t~ QED=2 QCD=2
output MyOutputDir
launch
\end{verbatim}
By default, \madgraph~5 imports the standard model. Therefore,  no specific command is needed for that.
As stated above, the process definition syntax is slightly different from the \madgraph~4 one. 
In  \madgraph~5,  spaces between particles names are mandatory.
As in \madgraph~4, it is possible to specify coupling orders.
If the coupling orders are not specified, then \madgraph~5 guesses
which interactions to allow based on the following rules:
\begin{itemize}
 \item If the orders defined in the model are \texttt{QED} and
   \texttt{QCD} only: The strong coupling is assumed to be dominant over the
   electroweak couplings, and the \texttt{QED} order is therefore set
   to its minimal possible value, while putting \texttt{QCD} to its
   maximal possible value. This provides the dominant contribution to
   the cross section, without the (negligible) sub-leading diagrams
   with additional QED couplings.
 \item If additional orders are present besides \texttt{QED} and
   \texttt{QCD}, then we allow any coupling order for any coupling,
   (if given couplings are preferred, the user needs to supply maximum
  orders in the process definition).
 \end{itemize}

The computation of the cross-section and the generation of partonic events 
is done with the command \texttt{launch}. Different options for cross
section calculation and event generation or matrix element evaluation can be found 
either with the \texttt{help launch} command or in the file MyOutputDir/README.

\subsection{Stop pair production}

This example shows how to evaluate the value of the square matrix element for a given point in phase-space. 
This is often used for testing the code and/or to interface \madgraph~with an external program. 

\begin{verbatim}
import model mssm
generate p p > t1 t1~
add process p p > t2 t2~
output standalone
launch
\end{verbatim}

First, the \texttt{mssm} model is imported. Then, the example shows
how to create multiple subprocesses.
The \texttt{generate} command clears all previously generated processes,
while the \texttt{add process} 
keeps all previous processes and adds the new process to the set.
If no output directory is given to the \texttt{output} command, the
output will be placed in an automatically named directory \texttt{PROC\_mssm\_0}.

\subsection{Slepton pair production}

In this example, we show how to  generate all slepton pair production
matrix elements for use in \pythia~8 (See Sec.~\ref{sec:pythia8}).
To generate the needed subprocesses, we could  use the \texttt{add process} command.
However, given that the number of sub-processes is quite large, this
is not very convenient. 
A much more handy solution is to use a  multi-particle label, similar to \texttt{p}/\texttt{j} for proton/jet:

\begin{verbatim}
import model mssm
define sl- = el- mul- ta1- er- mur- ta2-
define sl+ = el+ mul+ ta1+ er+ mur+ ta2+
generate p p > sl+ sl-
output pythia8
launch 
\end{verbatim}

In order to run event generation from the process, you will need to have \pythia~8 installed on your
computer. If the path to the \pythia~8 main directory is not given in
the \texttt{output} command, \madgraph~5 will look for it in the
default location (./pythia8). The default location can be modified by editing the
file \texttt{./input/mg5\_configuration.txt}.

\subsection{$W^+ jj$ production}

In this simple example we show how to create eps files containing the
Feynman diagrams for a set of processes:

\begin{verbatim}
generate p p > W+ j j
display diagrams ./
\end{verbatim}

The '.eps' files will be written in the output directory. Note that
unless a coupling order is specified, \madgraph~5 will minimize the
number of allowed \texttt{QED} orders, as explained in
Sec.~\ref{subsec:manual_tt} above; here \texttt{QED=1}. 
If you want the full expansion in  $\alpha_{em}$, you will need to specify where to stop the expansion, \ie, 
\begin{verbatim}
generate p p > W+ j j QED=3
display diagrams ./
\end{verbatim}

\subsection{Graviton-jet production}

In this example, we show how to evaluate the squared matrix element in
C++ for a more exotic model. We will use the specific case
of the graviton production with one additional jet in the
Randall-Sundrum model \cite{Randall:1999vf,Randall:1999ee}. The list
of command is the following:

\begin{verbatim}
import model RS
generate p p > y j
output standalone_cpp
launch
\end{verbatim}

In the model implementation, the name for the graviton is \texttt{y}. In order to know the name of all particles present in the model, you can use the command
 \texttt{display particles} and in order to get more information about a specific particle you can enter  \texttt{display particles y}.  
 In this examples they are three different coupling order labels
 \texttt{QED} / \texttt{QCD} / \texttt{QTD}; the latter is linked to the graviton sector of the theory.
Since they are three coupling orders, \madgraph~5 is not able to guess a suitable hierarchy between those order and set by default all orders to their maximal possible value.

\subsection{Gluino decay}

In this example we evaluate the partial decay width for gluino decay
into $u\bar u\chi_1^0$ through a left-handed squark:

\begin{verbatim}
generate go > ul > u~ u n1
output madevent
launch
\end{verbatim}

The particle(s) between the two $>$  are requested to be present in
all diagrams as $s$-channel propagators. Note that this condition does not imply that the
particle is strictly on-shell. Also note that the result of the decay width
calculation is given in GeV.

\subsection{Top-pair production with one leptonic decay}

In \madgraph~5, it is possible to specify the decay of a (nearly) on-shell particle (see Sec.~\ref{sec:decay}).
The syntax follows the logic of first stating the core process and then indicating the decay(s). 
If sub-decays are requested, they should be enclosed between parenthesis:
\begin{verbatim}
generate p p > t t~ QED=0, \
         (t > W+ b,   W+ > j j), \
         (t~ > w- b~, W- > l- vl~)
output madevent
launch
\end{verbatim}

The full process can be written on one line, or using the line
continuation symbol $\backslash$ to divide lines.

\section{The test suite}
\label{app:c}

\madgraph~5 features a test suite that allows to test the installation
as well as any further development of the code. During code
development, this test suite is extremely important in order to avoid
bugs, ensure the stability of the package and allow a robust
multi-developer approach. In this respect, it is considered mandatory
that the implementation of any new functionality is accompanied by
related tests. General advice is that the test suite should be as important and developed as the code itself and should be split into three different levels:

\begin{description}
\item[\underline{Unit tests:}] ~\\
Each part of the code (class or function) is tested by a series of
dedicated tests. The purpose is to fully check the behaviour of the
routine / class, \ie, check the output in some specific case, check
the behavior of the code in case of wrong input, etc. Currently \madgraph~5 includes more than 400 independent unittests. 
\item[\underline{Acceptance tests:}] ~\\
This part simulates the instructions entered by the user and checks that all modules are correctly  interfaced. This corresponds to check that a series of examples are correctly running and provide the expected results.
\item[\underline{Parallel tests:}] ~\\
These tests check that the output of \madgraph~5 are the same of those obtained by \madgraph~4. These checks are made for different models (SM/ MSSM/ HEFT) and using both 
\ufo\  and version 4 models.
\end{description}

In order to efficiently run these tests, we have implement a script
which automatically detects all the tests available. The
tests can be filtered by name or file, in order to run only a subset
of tests. A similar 
test module is now present by default in
\python~2.7. However, since \madgraph~5 is by design also compatible
with \python~2.6, a dedicated test code is included in the distribution.  In order to run the different test suites, the user can type (respectively for unittest / acceptance tests / parallel tests):

\begin{verbatim}
./tests/tests_manager.py
./tests/tests_manager.py -p A
./tests/tests_manager.py -p P
\end{verbatim}

 \bibliographystyle{jhep}
\bibliography{physics}

\end{document}